\documentclass[]{input/pasj02}%\draft
\usepackage[switch,mathlines]{lineno} % add line number to manuscript
\usepackage{xcolor}
\jyear{2024}
\Received{}%{yyyy/mm/dd}
\Accepted{}%{yyyy/mm/dd}
\begin{document} 
\title{{Circum-nuclear eccentric gas flow} in the Galactic Center revealed by ALMA CMZ Exploration Survey (ACES)} 
%%%%%%%%%%%%%%%%%%%%%%  
%%%%%%%%%%% ACES Author list %%%%%%%%%%%%%%%%%
\def\afmark{\altaffilmark}
\def\aftext{\altaffiltext} 
\def\scr{\scriptsize}  
\author{
%%%% Authors: Please edit below %%%%%%
Yoshiaki \textsc{Sofue}\afmark{1}\afmark{*}\orcid{0000-0002-4268-6499}\email{sofue@ioa.s.u-tokyo.ac.jp}\aftext{1}{\scr Institute of Astronomy, The University of Tokyo, Mitaka, Tokyo 181-0015, Japan}, 
%sofue@ioa.s.u-tokyo.ac.jp 
%
Tomoharu \textsc{Oka}\afmark{2}\orcid{0000-0002-5566-0634}\aftext{2}{\scr Department of Physics, Faculty of Science and Technology, Keio University, 3-14-1 Hiyoshi, Yokohama, Kanagawa 223-8522, Japan}, 
%tomo@phys.keio.ac.jp
%
%%%%%%%%%%%%%%%ACES PI, CoI, WP Leads %%%%%%%%%%%%
\orcid{0000-0001-6353-0170}{Steven N. \textsc{Longmore}}\afmark{3,4}\aftext{3}{Astrophysics Research Institute, Liverpool John Moores University, IC2, Liverpool Science Park, 146 Brownlow Hill, Liverpool L3 5RF, UK}\aftext{4}{Cosmic Origins Of Life (COOL) Research DAO, https://coolresearch.io}
% s.n.longmore@ljmu.ac.uk
%
\orcid{0000-0001-7330-8856}{Daniel \textsc{Walker}}\afmark{5}\aftext{5}{{\scr UK ALMA Regional Centre Node, Jodrell Bank Centre for Astrophysics, Oxford Road, The University of Manchester, Manchester M13 9PL, United Kingdom}},
%danielwalker274@gmail.com
%
\orcid{0000-0001-6431-9633}{Adam \textsc{Ginsburg}}\afmark{6},\aftext{6}{Department of Astronomy, University of Florida, P.O. Box 112055, Gainesville, FL 32611}
%adam.g.ginsburg@gmail.com
%
\orcid{0000-0001-9656-7682}{Jonathan D. \textsc{Henshaw}}\afmark{3,7},\aftext{7}{Max Planck Institute for Astronomy, K\"{o}nigstuhl 17, D-69117 Heidelberg, Germany}
%jonathan.d.henshaw@gmail.com
%
\orcid{0000-0001-8135-6612}{John \textsc{Bally}}\afmark{8},\aftext{8}{Center for Astrophysics and Space Astronomy; Department of Astrophysical and Planetary Sciences; University of Colorado, Boulder, CO 80389, USA}
%john.bally@colorado.edu
%
\orcid{0000-0003-0410-4504}{Ashley T. \textsc{Barnes}}\afmark{9},\aftext{9}{European Southern Observatory (ESO), Karl-Schwarzschild-Strasse 2, D-85748 Garching, Germany}
%ashleybarnes.astro@gmail.com
%
\orcid{0000-0002-6073-9320}{Cara \textsc{Battersby}}\afmark{10},\aftext{10}{Department of Physics, University of Connecticut, 196A Auditorium Road, Unit 3046, Storrs, CT 06269, USA}
%cara.battersby@uconn.edu
%
\orcid{0000-0001-8064-6394}{Laura \textsc{Colzi}}\afmark{11},\aftext{11}{Centro de Astrobiología (CAB), CSIC-INTA, Carretera de Ajalvir km 4, Torrejón de Ardoz, 28850 Madrid, Spain}
% lcolzi@cab.inta-csic.es
%
Paul \textsc{Ho}\afmark{12},\aftext{12}{AS/NTU Astronomy-Mathematics Building,  Roosevelt Rd, Taipei 10617, Taiwan}\orcid{0000-0002-3412-4306}
Izaskun \textsc{Jimenez}-\textsc{Serra}\afmark{11}%\aftext{WP3}{WP3 CO PI}
\orcid{0000-0003-4493-8714},
%ijimenez@cab.inta-csic.es
%
\orcid{0000-0002-8804-0212}J.~M.~Diederik \textsc{Kruijssen}\afmark{4},
%kruijssen@coolresearch.io
%
Elizabeth \textsc{Mills}\afmark{13},\aftext{13}{Department of Physics and Astronomy, University of Kansas, 1251 Wescoe Hall Drive, Lawrence, KS 66045, USA}\orcid{0000-0001-8782-1992}
\orcid{0000-0002-6362-8159}Maya A. \textsc{Petkova}\afmark{14},\aftext{14}{Space, Earth and Environment Department, Chalmers University of Technology, SE-412 96 Gothenburg, Sweden}
%maya.petkova28@gmail.com
%
\orcid{0000-0001-6113-6241}{Mattia C. \textsc{Sormani}}\afmark{15},\aftext{15}{Como Lake centre for AstroPhysics (CLAP), DiSAT, Universit{\`a} dell’Insubria, via Valleggio 11, 22100 Como, Italy }
%\aftext{16}{Department of Physics, University of Surrey, Guildford GU2 7XH, UK}
%mattia.sormani@uninsubria.it
%
\orcid{0009-0002-7459-4174}{Jennifer \textsc{Wallace}}\afmark{10},\aftext{10}{Department of Physics, University of Connecticut, 196A Auditorium Road, Unit 3046, Storrs, CT 06269, USA}
%jennifer.2.wallace@uconn.edu
%
%%%%%%%%%%% ACES members  et al %%%%%%%%%%%%%% 
\orcid{0000-0003-3341-6144}{Jairo  \textsc{Armijos-Abenda\~no}}\afmark{17},\aftext{17}{Observatorio Astron\'omico de Quito, Observatorio Nacional, Escuela Polit\'ecnica Nacional, Interior del Parque La Alameda, 170136, Quito, Ecuador}
%jarmijos090@gmail.com
%
\orcid{0000-0003-0980-6871}{Katarzyna M. \textsc{Dutkowska}}\afmark{18},\aftext{18}{Leiden Observatory, Leiden University, P.O. Box 9513, 2300 RA Leiden, The Netherlands}
%dutkowska@strw.leidenuniv.nl
%
\orcid{0000-0001-0000-0000}{Rei \textsc{Enokiya}}\afmark{19}, \aftext{19}{National Astronomical Observatory of Japan, 2-21-1 Osawa, Mitaka, Tokyo 181-8588, Japan}
% rei.enokiya@nao.ac.jp
%
%\orcid{0000-0002-8966-9856}{Yasuo \textsc{Fukui}}\afmark{20},\aftext{20}{Department of Physics, Nagoya University, Chikusa-ku, Nagoya 464-8602, Japan}
%
Pablo \textsc{Garc\'ia}\afmark{21,22},\aftext{21}{Chinese Academy of Sciences South America Center for Astronomy, National Astronomical Observatories, CAS, Beijing 100101, China}\aftext{22}{Instituto de Astronom\'ia, Universidad Cat\'olica del Norte, Av. Angamos 0610, Antofagasta, Chile}
%pablo.garcia@nao.cas.cn
%
\orcid{0000-0002-1313-429X}{Savannah \textsc{Gramze}}\afmark{6},
%savannahgramze@ufl.edu
%
%Andres \textsc{Guzman}\afmark{23},\aftext{23}{Joint ALMA Observatory, Alonso de Cordova 3107, Vitacura 763-0355, Santiago de Chile, Chile}
%
\orcid{0000-0002-7495-4005}{Christian \textsc{Henkel}}\afmark{24},\aftext{24}{MPIfR, Auf dem H\"ugel 69, Bonn, Germany}
%chenkel@mpifr-bonn.mpg.de
\orcid{0000-0001-9155-3978}{Pei-Ying \textsc{Hsieh}}\afmark{19},
%\aftext{19}{National Astronomical Observatory of Japan, 2-21-1 Osawa, Mitaka, Tokyo 181-8588, Japan}
%
Yue \textsc{Hu}\afmark{25},\aftext{25}{Institute for Advanced Study, 1 Einstein Drive, Princeton, NJ 08540, USA}\orcid{0000-0002-8455-0805}
% yuehu@ias.edu
\orcid{0000-0003-4140-5138}{Katharina \textsc{Immer}}\afmark{9},
\orcid{0000-0002-9255-4742}{Yuhei \textsc{Iwata}}\afmark{26,27},\aftext{26}{Mizusawa VLBI Observatory, National Astronomical Observatory of Japan, 2-12 Hoshigaoka, Mizusawa, Oshu, Iwate 023-0861, Japan}\aftext{27}{Astronomical Science Program, Graduate Institute for Advanced Studies, SOKENDAI, 2-21-1 Osawa, Mitaka, Tokyo, 181-8588, Japan}
%yuhei.iwata@nao.ac.jp
%
%\orcid{0000-0003-0416-4830}{Desmond \textsc{Jeff}}\afmark{6,28},\aftext{28}{National Radio Astronomy Observatory, 520 Edgemont Road, Charlottesville, VA 22903, USA}
%%
Janik \textsc{Karoly}\afmark{28},\aftext{28}{Department of Physics and Astronomy, University College London, Gower Street, London WC1E 6BT, UK}\orcid{0000-0001-5996-3600}
%j.karoly@ucl.ac.uk
%
Ralf S.\ \textsc{Klessen}\afmark{29,30,31,32},
\aftext{29}{Universit\"{a}t Heidelberg, Zentrum f\"{u}r Astronomie, Institut f\"{u}r Theoretische Astrophysik, Albert-Ueberle-Str.\ 2, 69120 Heidelberg, Germany}
\aftext{30}{Universit\"{a}t Heidelberg, Interdisziplin\"{a}res Zentrum f\"{u}r Wissenschaftliches Rechnen, Im Neuenheimer Feld 225, 69120 Heidelberg, Germany}
\aftext{31}{Center for Astrophysics $\vert$ Harvard \& Smithsonian, 60 Garden Street, Cambridge, MA, 02138, USA}
\aftext{32}{Elizabeth S. and Richard M. Cashin Fellow at the Radcliffe Institute for Advanced Studies at Harvard University, 10 Garden Street, Cambridge, MA 02138, USA}\orcid{0000-0002-0560-3172}
%cu120@uni-heidelberg.de
%klessen@structures.uni-heidelberg.de
%
Kotaro \textsc{Kohno}\afmark{1},\orcid{0000-0002-4052-2394}
%kkohno@ioa.s.u-tokyo.ac.jp
%
Mark R. \textsc{Krumholz}\afmark{33},\aftext{33}{Research School of Astronomy and Astrophysics, Australian National University, Cotter Road, Weston ACT 2611, Australia}\orcid{0000-0003-3893-854X}
\orcid{0000-0002-5776-9473}{Dani \textsc{Lipman}}\afmark{10}, 
% dani.lipman@uconn.edu
%
Mark R. \textsc{Morris}\afmark{34},\aftext{34}{Department of Physics and Astronomy, University of California, Los Angeles, CA 90095, USA} \orcid{0000-0002-6753-2066} 
\orcid{0000-0002-6379-7593}{Francisco \textsc{Nogueras-Lara}}\afmark{9}, 
%% francisco.nogueraslara@eso.org
%
%\orcid{0000-0000-0000-0000}Mairi \textsc{Nonhebel}\afmark{9,35},\aftext{35}{SUPA,School of Physics and Astronomy, University of St. Andrews, North Haugh, St. Andrews KY16 9SS, UK}
%
%\orcid{0000-0001-8224-1956}{J\"urgen \textsc{Ott}}\afmark{36},\aftext{36}{National Radio Astronomy Observatory, P.O. Box O, 1011 Lopezville Road, Socorro, NM 87801, USA} 
% 
\orcid{0000-0002-3972-1978}{Jaime E. \textsc{Pineda}}\afmark{37},\aftext{37}{MPI for Extraterrestrial Physics, Giessenbachstr. 1, D-85748, Garching, Germany}
%jaime.e.pineda@gmail.com
%
\orcid{0000-0001-9281-2919}{Sergio \textsc{Mart\'in}}\afmark{38,39},\aftext{38}{European Southern Observatory, Alonso de C\'ordova, 3107, Vitacura, Santiago 763-0355, Chile}\aftext{39}{Joint ALMA Observatory, Alonso de C\'ordova, 3107, Vitacura, Santiago 763-0355, Chile}
%smartin@eso.org
%
\orcid{0009-0009-5346-7329}{Miguel Angel \textsc{Requena-Torres}}\afmark{40},\aftext{40}{Department of Physics, Astronomy, and Geosciences, Towson University, Towson, MD 21252, USA} 
%ma.requena.torres@gmail.com
%
\orcid{0000-0002-2887-5859}{Víctor M. \textsc{Rivilla}}\afmark{11}, 
%vrivilla@cab.inta-csic.es
%
\orcid{0000-0001-5389-0535}{Denise \textsc{Riquelme-V\'asquez}}\afmark{41},\aftext{41}{Departamento de Astronom\'ia, Universidad de La Serena, Ra\'ul Bitr\'an 1305, La Serena, Chile}
%denise.riquelme.vasquez@gmail.com
%
\orcid{0000-0002-3078-9482}{\'Alvaro S\'anchez-Monge}\afmark{42,43},\aftext{42}{Institut de Ci\`encies de l'Espai (ICE), CSIC, Campus UAB, Carrer de Can Magrans s/n, E-08193, Bellaterra (Barcelona), Spain}\aftext{43}{Institut d'Estudis Espacials de Catalunya (IEEC), E-08860, Castelldefels (Barcelona), Spain}
%%asanchez@ice.csic.es
%
\orcid{0000-0002-3941-0360}{Miriam \textsc{G. Santa-Maria}}\afmark{6},
%miriam.garcias01@gmail.com
Howard A.~\textsc{Smith}\afmark{31}
%\aftext{31}{Center for Astrophysics $\vert$ Harvard \& Smithsonian, 60 Garden Street, Cambridge, MA, 02138, USA},
%hsmith@cfa.harvard.edu
%
%Tabassum S \textsc{Tanvir}\afmark{44},\aftext{44}{Department of Physics and Astronomy, Iowa State University, 2323 Osborn Drive, Ames, IA 50010, USA} \orcid{0000-0002-0862-0701}
%  
\orcid{0000-0003-1841-2241}Volker \textsc{Tolls}\afmark{31},
%vtolls@cfa.harvard.edu
%
and
\orcid{0000-0002-9279-4041}{Q. Daniel \textsc{Wang}}\afmark{45}\aftext{45}{Department of Astronomy, University of Massachusetts, Amherst, MA 01003, USA}
%wqd@umass.edu 
%%%%%%%%%%%%%%%%%%%%%%%%%%%%%%%%%%%%%%
}  % <=DON'T TOUCH
%%%%%% + shared editors overleaf : PLEASE ADD YOU TO THE AUTHOR LIST%%%%%%%
%rojitabuddhacharya@gmail.com
%nbutterf@nrao.edu
%grant.tremblay@cfa.harvard.edu
%sandyzeng.28@gmail.com
%christoph.federrath@anu.edu.au
%ajschmie@nrao.edu

%% List of Key Words: https://academic.oup.com/pasj/pages/Pasj_Keywords 
\KeyWords{Galaxy : center --- Galaxy : structure --- ISM : clouds --- ISM : molecules --- ISM : kinematics and dynamics}  
\maketitle
%%%%%%%%%%%%%%%%%%%%%%%%%%%%%%%%
 
\def\be{\begin{equation}}\def\ee{\end{equation}}\def\vlsr{v_{\rm LSR}} \def\Vlsr{\vlsr}\def\Msun{M_\odot} \def\vr{v_{\rm r}}\def\deg{^\circ}  \def\d{^\circ}
\def\vrot{V_{\rm rot}} \def\Vrot{\vrot}\def\co{$^{12}$CO } \def\coth{$^{13}$CO $(J=1-0)$} \def\Xco{X_{\rm {^{12}}CO}}   \def\Tb{T_{\rm B}} \def\Tp{T_{\rm p}}\def\Htwo{H$_2$} \def\htwo{H$_2$}  \def\Kkms{K km s$^{-1}}  \def\Hcc{{\rm H \cm^{-3}}} \def\kms{km s$^{-1}$} \def\Ico{I_{\rm CO}}  \def\Kkms{K \kms }\def\mH{m_{\rm H}}  \def\Ico{I_{\rm ^{12}CO}} \def\Icoth{I_{\rm ^{13}CO}} 
 \def\htwo{H$_2$} \def\Tb{T_{\rm B}}   \def\mH{m_{\rm H}} \def\ekms{{\rm \ km \ s^{-1}}}  \def\epc{{\rm \ pc} }  \def\Hii{HII} \def\apj{ApJ} \def\aap{A\&A} \def\mnras{MNRAS} \def\pasj{PASJ} \def\aj{AJ} \def\xcounit{H$_2$ cm $^{-2}$ [K km s$^{-1}]^{-1}$}  \def\log{{\rm log}}  \def\tc{t_{\rm C}} \def\fc{f_{\rm C}} \def\SFR{{\rm SFR}} \def\sfr{{\rm SFR}}\def\tc{t_{\rm c}}\def\lc{l_{\rm c}}\def\vc{v_{\rm c}}\def\tp{t_{\rm p}}\def\rc{r_{\rm c}}\def\nc{n_{\rm c}}\def\pcc{p_{\rm c}}  
\def\Msun{M_{\odot \hskip-4.8pt \bullet}}      
\def\kms{km s$^{-1}$}  \def\deg{^\circ}   \def\Htwo{H$_2$\ }  \def\fmol{f_{\rm mol}} \def\Fmol{ $f_{\rm mol}$ }  \def\sfu{\Msun~{\rm y^{-1}~kpc^{-2}}} \def\sfuvol{\Msun~{\rm y^{-1}~kpc^{-3}}}\def\log{{\rm log}}
\def\hcc{{\rm H~cm^{-3}}} \def\Hcc{ $\hcc$ }\def\Htot{ H$_{\rm tot}$ } \def\ssfr{\Sigma_{\rm SFR}} \def\vsfr{\rho_{\rm SFR}} \def\sfr{{\rm SFR}}\def\H{{\rm H}}\def\cm{{\rm cm}}\def\kpc{{\rm kpc}} \def\bc{\begin{center}}\def\ec{\end{center}} 
\def\xcounit{\Htwo cm$^{-2}$ [K \kms]} \def\pc{{\rm pc}}\def\My{{\rm My}} \def\kpc{{\rm kpc}}\def\rc{r_{\rm c}} \def\vc{v_{\rm c}}
\def\urho{\Msun \pc^{-3}}\def\urhohtwo{{\rm H_2} \cm^{-3}} \def\nc{n_{\rm c}} 
\def\vexpa{v_{\rm expa}} \def\rbub{r_{\rm b}}\def\x{\times}\def\xfour{\times 10^4}\def\xthree{\times 10^3}\def\xfive{\times 10^5}\def\xfifty{\times 10^{50}}\def\xmtwe{\times 10^{-20}} \def\sigv{\sigma_v} \def\Rbow{R_{\rm bow}} \def\Rzero{R_0} \def\Lcone{L_{\rm cone}} \def\Lsun{L_\odot}\def\Rhii{R_{\rm HII}} \def\nuv{n_{\rm UV}} \def\ni{n_{\rm i}} \def\ne{n_{\rm e}}\def\ar{a_{\rm r}}
\def\Te{T_{\rm e}}\def\Tn{T_{\rm n}}\def\cosh{{\rm cosh}}\def\({\left(} \def\){\right)}\def\[{\left[} \def\]{\right]}\def\Hcc{H cm$^{-3}} \def\Hsqcm{H cm$^{-2}$} \def\L{\mathcal{L}}\def\Rc{R_{\rm c}} \def\rhom{\rho_{\rm m}} \def\rhoc{\rho_{\rm c}} \def\V{V_{\rm rot}} \def\Vp{V_{\rm pat}} \def\vpat{V_{\rm pattern}}
\def\red{\textcolor{red}} \def\blue{\textcolor{blue}} 
\def\ss{\subsection}\def\sss{\subsubsection}
\def\hcnaces{H$^{13}$CN $(J=1-0)$}
\def\hcn{H$^{13}$CN $(J=1-0)$}
\def\cs{CS ($J=2-1$)}
\def\csaces{CS ($J=2-1$)}
\def\hcnaste{HCN ($J=4-3$)}
\def\dvdl{d\vlsr/dl}
\def\sgrastar{Sgr A$^*$} 
\def\vex{V_{\rm ex}}
\def\Jybeam{Jy beam$^{-1}$}
\def\cc{cm$^{-3}$}
\def\asec{''\!\!}
\def\degd{^\circ\!\!}
\def\htwocc{\htwo cm$^{-3}$ }
\def\G02{G+0.02-0.02+100}
\def\vex{V_{\rm expa}}
\def\vexpa{V_{\rm expa}}
\def\ekpc{{\rm kpc}}
 \def\Htwocc{\Htwo cm$^{-3}$}
 \def\degp{\deg\!.}
 \def\Jyb{Jy beam$^{-1}$}  
%%%%%%%%%%%%%%%%%%%%%%%%%%%%%%%
\begin{abstract} 
We analyze the \cs\ line cube from the internal data release obtained by the large-scale program "ALMA CMZ Exploration Survey (ACES)" to investigate the kinematic structure of the innermost $\sim 10$ pc region of the Galaxy, which contains the high-velocity compact cloud (HVCC) at $(l,b,\vlsr)\sim(+0\degp02,-0\degp02, 100 \ekms)$ (hereafter G0.02). 
The longitude-velocity diagram (LVD) of the cloud draws an elliptical structure, which is interpreted as an orbital trajectory in the $(l,\vlsr)$ space of a noncircular (eccentric) motion of the molecular gas in the gravitational potential of an extended mass distribution in the central 10 pc of the Galaxy.  
We argue that G0.02 is a kinematic tracer of the inner potential, a rare case of a dense gas following an eccentric orbit in the nuclear gravitational field. 
\end{abstract} 
%\pagewiselinenumbers  
 %%%%%%%%%%%%%%%%%%%%%%
\section{Introduction}
\label{intro}   
In our recent study of the molecular gas distribution in the {central molecular zone (CMZ) \citep{PaperI})}, we have identified six arms named "Galactic-Center Arms (GCAs) I to VI" by analyzing the molecular-line cube data of the \coth\ and \hcnaste\ lines from single dish observations, and \csaces\ and \hcnaces\ lines from ALMA (Atacama Large Millimeter/submillimeter Array) taken by the large project ACES (ALMA CMZ Exploration Survey) (S. Longmore et al. in preparation). GCA I and II compose the 120 pc molecular ring that shares about 90\% of the CMZ mass, Arm III to V are intermediate arms, and Arm VI is the innermost molecular arm of radius $\sim 3$ pc known as the circumnuclear disc (CND).
We have tried to understand these arms under a unified view of Galactic arms and rings rotating in the common gravitational potential around \sgrastar.

In addition to such "ordinary" arms represented by straight ridges in the longitude-velocity diagrams (LVD), there are LV features that exhibit high-velocity noncircular motions, which cannot be attributed to the arms in ordinary Galactic rotation.
The CND of radius $\sim 3$ pc is well known for exhibiting such peculiar motions ALMA \citep{2021ApJ...913...94H}.

The high-velocity compact cloud (HVCC; hereafter \G02, or G0.02) located at $(l,b, \vlsr)\simeq (0\degp02,-0\degp02,+100 \ekms)$ at a projected distance of 10 pc from \sgrastar\ is the most typical example of such clouds showing anomalous kinematics \citep{oka+1999,2008PASJ...60..429O,2022ApJS..261...13O,2023ApJ...950...25I}. 
G0.02 is outside the CND and was discovered by CO-line observations using the Nobeyama 45 m telescope and is named CO 0.02-0.02 \citep{oka+1999}.
It has a size of $\sim 3$ pc, a velocity width as large as $\sim 50$ \kms, a molecular mass of $\sim 10^5\Msun$ inferred from the luminosity of the \co\ line. 
It is suggested that the origin of the extreme physical condition is due to a strong disturbance due to a nearby supernova explosion \citep{2008PASJ...60..429O,2023ApJ...950...25I}.

In this paper, we examine the kinematics of the LV features of G0.02\ and related structures from the point of view of galactic dynamics by analyzing the 3D cube of the \csaces\;line emission observed with ALMA by ACES. 
We adopted a distance $R_0=8.2$ kpc to the GC, close to the recent measurement
\citep{gravity+2019}. 

%%%%%%%%%%%%%%%%%%%%%%%%%%%%%%%%%%%%
\section{Longitude-velocity ellipse}
\ss{Data}  
The molecular line cubes in this work were taken from the internal release version of the 12m + 7m + TP (Total Power) mode data from the ALMA Cycle 8 Large Program "ALMA Central Molecular Zone Exploration Survey" (ACES, 2021.1.00172.L; Longmore et al. in preparation). ACES observed the CMZ in ALMA Band 3, covering a frequency range of $\sim$86--101~GHz across six spectral windows of varying spectral resolution and bandwidth.
See Paper I \citep{PaperI} for the data used in this paper.  

In this work, we use cubes in the \cs\;line at a frequency of 97.9810 GHz with an angular resolution of $2\asec.21$ and an rms noise of 0.0038 \Jybeam (0.10 K) with velocity channels of 1.45 \kms.
We also have the \hcn\ line at 86.3399 GHz, 
but the velocity coverage ($-200\le \vlsr\le 200$ \kms) in the current internal release data was not sufficient for the purpose of this paper to explore G0.02.
So, we restricted our analysis to the \csaces\ line that covers $-220\le\vlsr\le 220$ \kms.
We note that a preliminary analysis in \hcnaces\;revealed almost the same kinematical properties of G0.02 as those of the \csaces\ line.

The intensity scales are in \Jybeam\ (1 \Jybeam$=26.1$ K in brightness temperature at 98 GHz). 
The used \csaces\ cube covers the CMZ at $-0\degd.6 \lesssim l \lesssim +0\degd.9$ and $-0\degd.3 \lesssim b \lesssim +0\degd.1$ with spatial and velocity grids of ($0\asec.5 \times 0\asec.5\times 0.15$ \kms).
In the present work, we cut out a region at $-0\degd.25 \le l \le +0\degd.15$ and $-0\degd.1 \le b \le +0\deg$ for a detailed analysis of the circum-nuclear region centered on \sgrastar. 

\begin{figure}   
\begin{center}   
(A)\includegraphics[width=6cm] {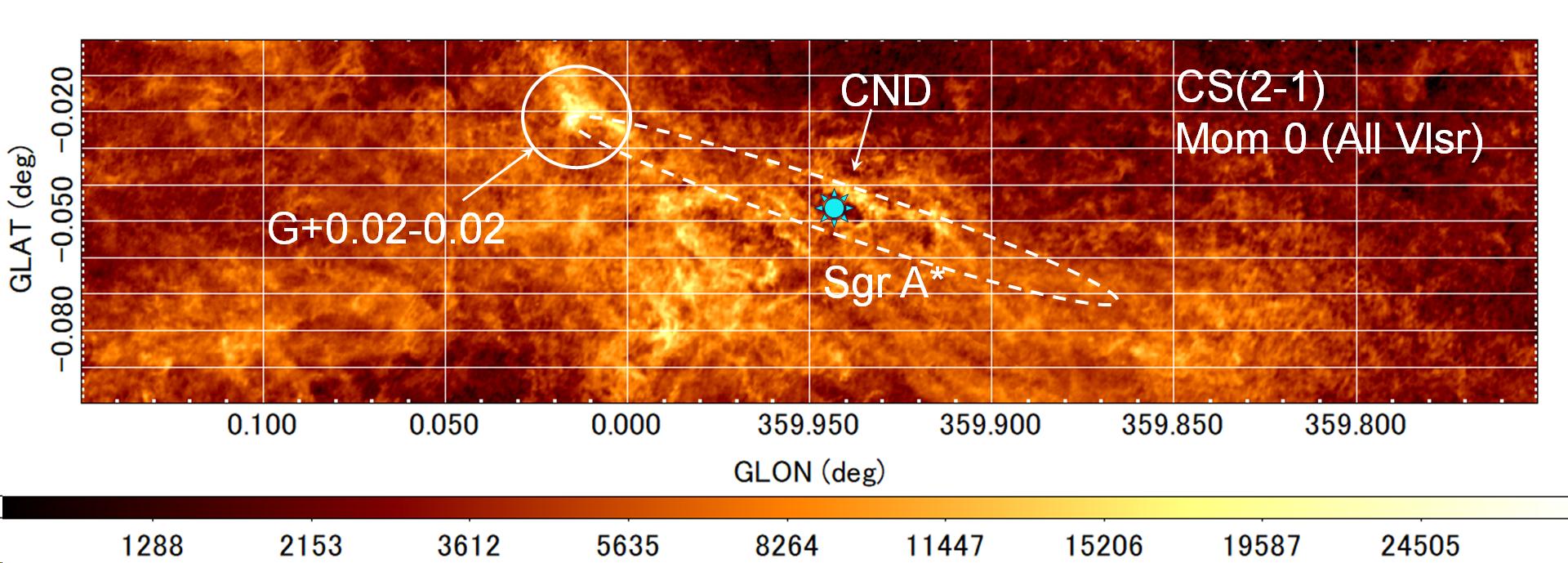}   \\
(B)\includegraphics[width=6cm]{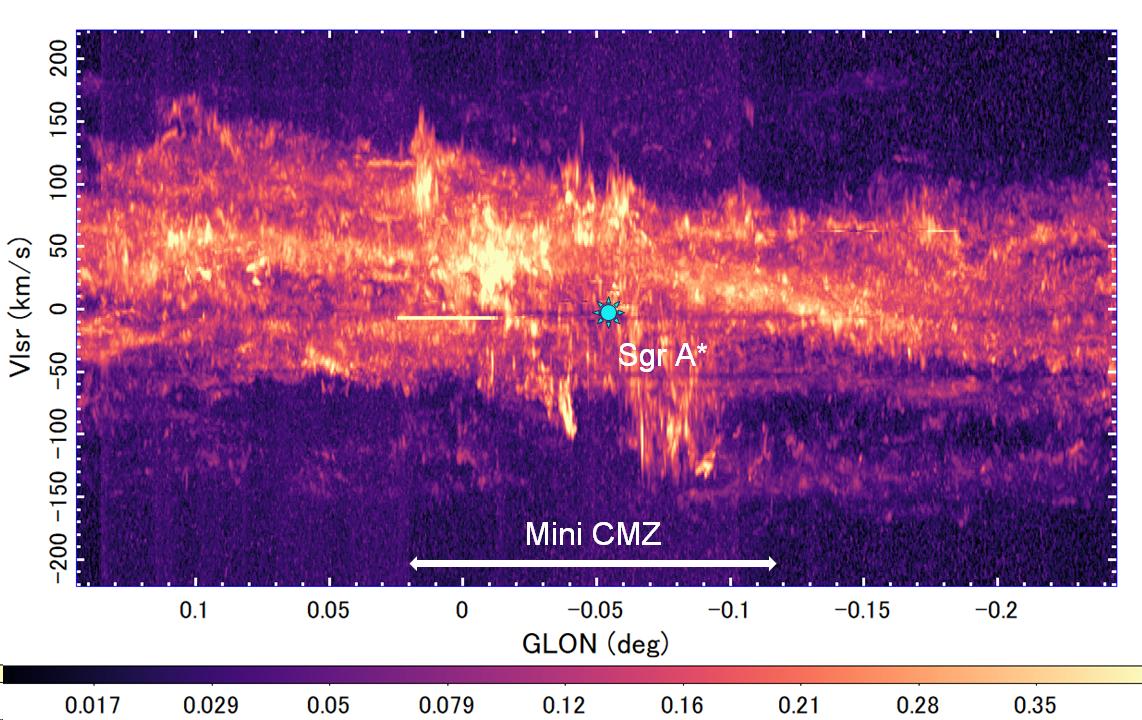}  \\  
(C)\includegraphics[width=6cm]{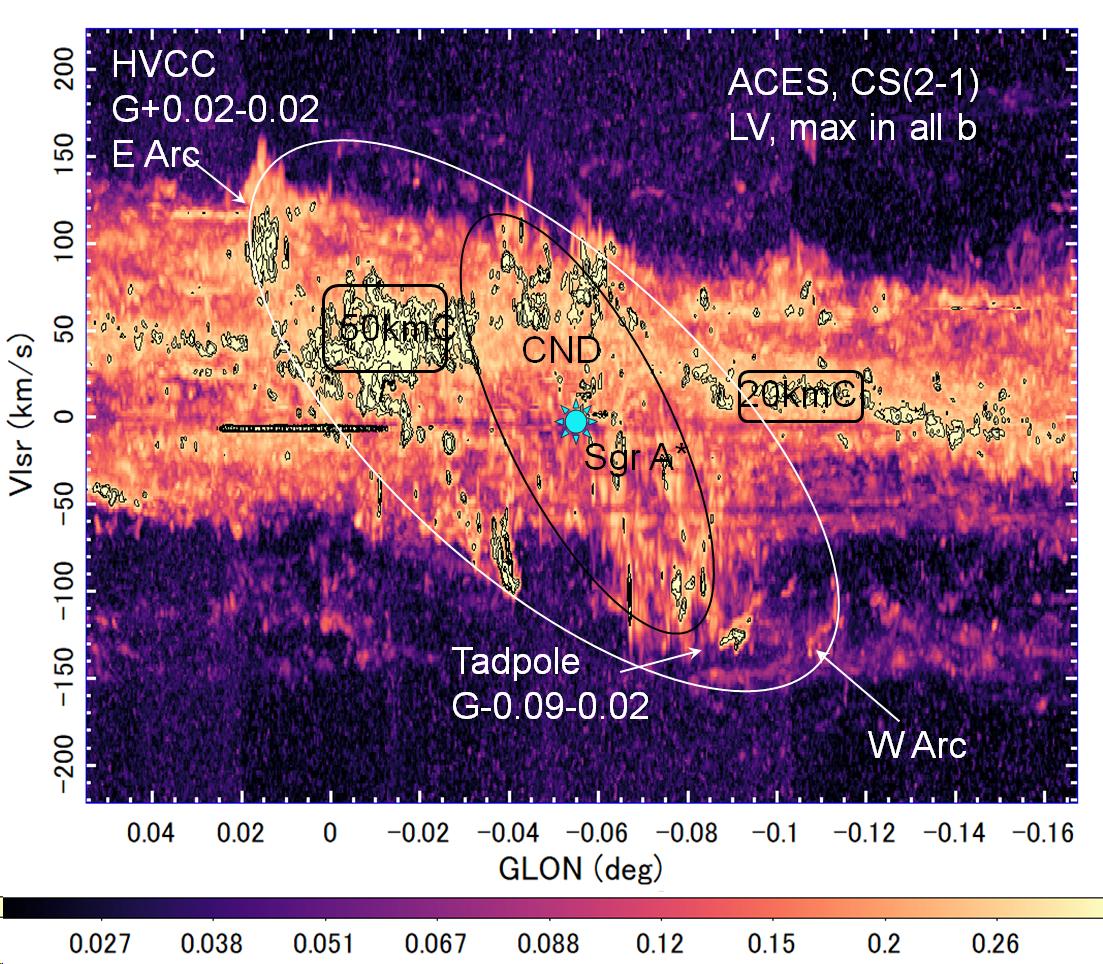} \\ 
(D)\includegraphics[width=6.2cm]{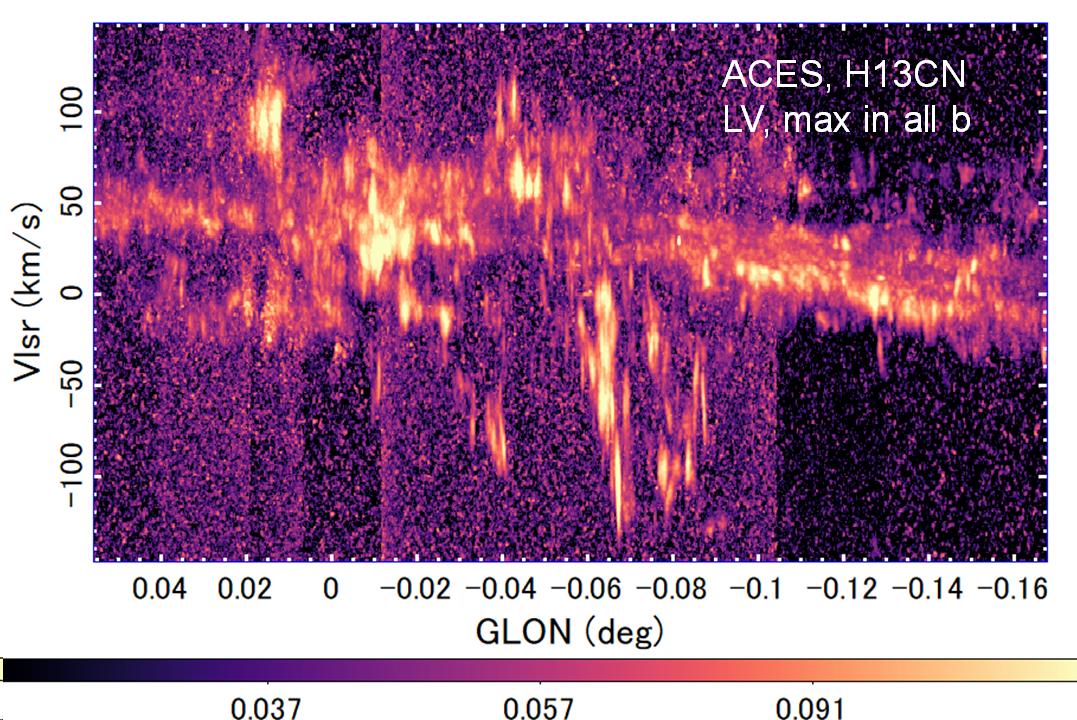} 
\end{center} 
\caption{
[A] ACES Moment 0 map of the \cs\ line in \Jyb m s$^{-1}$.
The circle indicates G0.02 and the dashed ellipse outlines an inclined 10-pc ring.
[B] LVD of the maximum intensity in \Jyb.
Note the high-velocity structures at $|l-l_{\rm Sgr \ A^*}|\lesssim 0\degp07$ (10 pc), which we call the 'mini CMZ'.  
[C] Same, but enlarged. 
The LV ellipse and some well known objects are indicated.  
{[D] Same, but in the \hcnaces\ line.}
The data were taken from the internal release of ACES (Longmore et al. in preparation). 
{Alt text: Moment 0 map and LVDs in the central 10 pc region of the GC in the \cs\ line from ACES. }
}
\label{mini-cmz}  
\end{figure}

\begin{figure}  
\begin{center}     
\hskip -5mm Obs. \cs \hskip 2cm Model\\
\includegraphics[height=4cm]{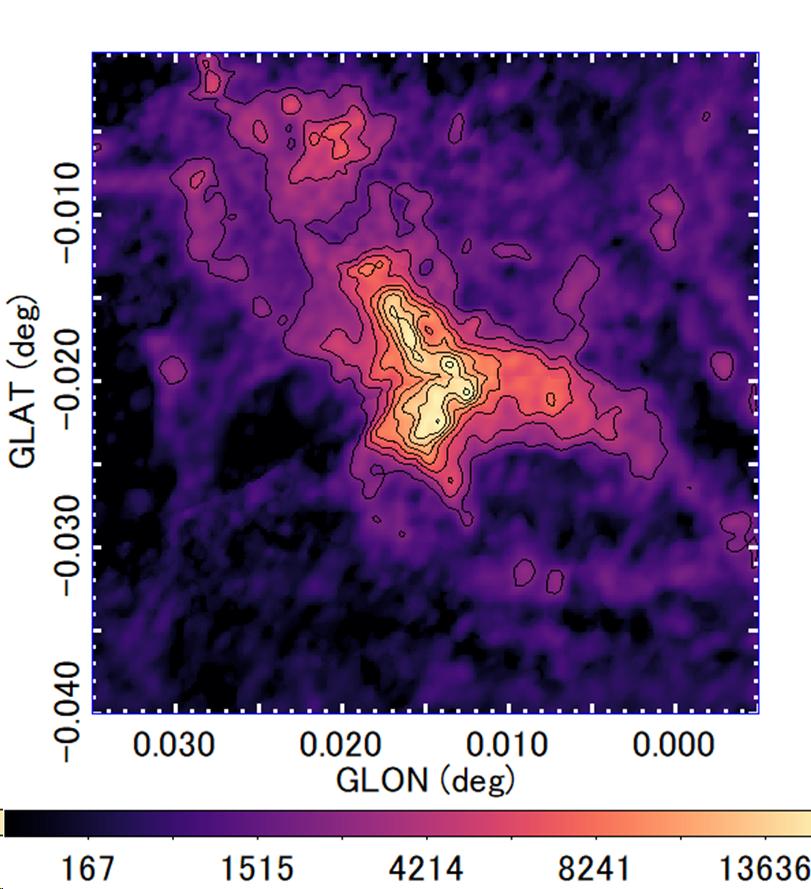}
\includegraphics[height=4cm]{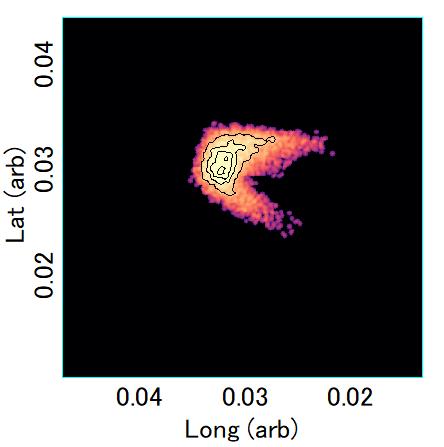}  \\
\includegraphics[height=4cm]{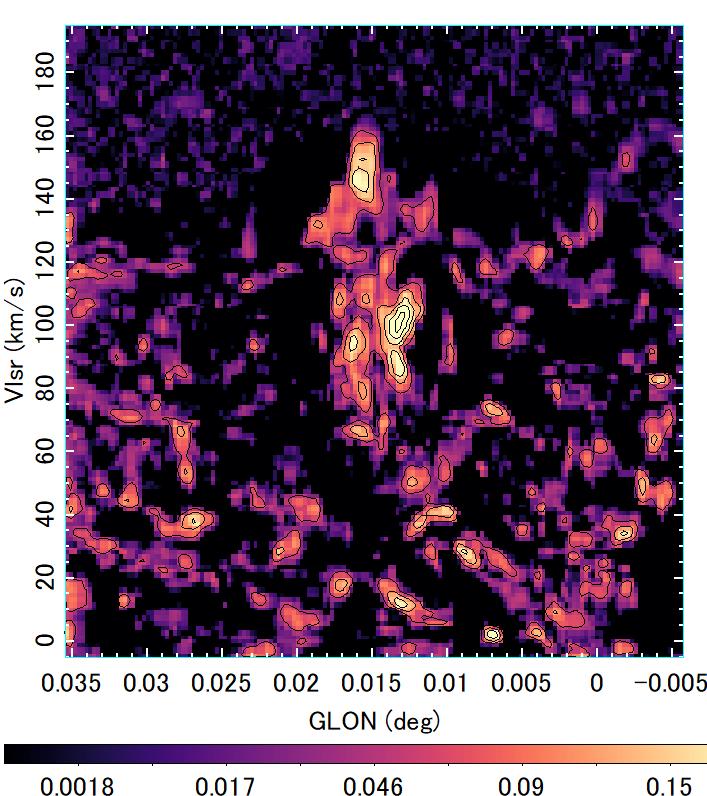}  
\includegraphics[height=4cm]{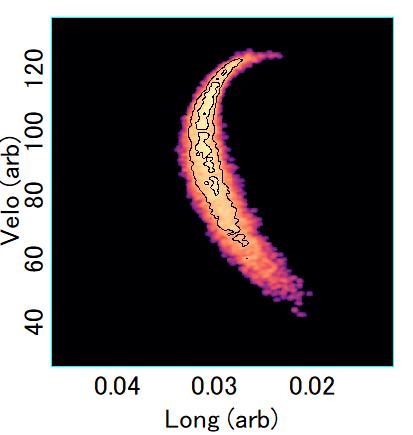} 
\end{center} 
\caption{[Top left] Moment 0 map of G0.02 in the \csaces\ line (Jy/b m s$^{-1}$) at $\vlsr\ge 75$ \kms. Contours are every 2 Jy beam$^{-1}$ m s$^{-1}$. 
[Bottom left] LVD of maximum intensity, where horizontally extended components have been subtracted.  Contours are every 0.05 Jy beam$^{-1}$. 
[Right panels] Simulation of G0.02 for a similar condition as F in figure \ref{models}. 
{The general property is well reproduced except for the high velocity straight LV wing.}
{Alt text: Comparison of simulated moment 0 map and LVD with the observation. }}
\label{G0.02simu}  
\end{figure}

\ss{Mini-CMZ and G0.02} 
Figure \ref{mini-cmz} (panel A) shows a moment 0 map of the \cs\ line integrated from $-220$ to $+220$ \kms.
Figure \ref{G0.02simu} enlarges the region around G0.02, which appears as a compact triangular cloud composed of several bright knots and arcs at $(l,b)\sim (0\deg\!\!.02, -0\degp02)$. 
It also exhibits an arc-shaped cavity concave to G0.005-0.028, as reported by \citet{2008PASJ...60..429O} .

In panel B of figure \ref{mini-cmz}, we show a longitude-velocity diagram (LVD) of the maximum intensity along latitude at each Galactic longitude in the \csaces\ line between $l=-0\degp25$ and $+0\degp15$.
Comparison of this figure with the LVD of the Galactic plane survey in the \co line \citep{2001ApJ...547..792D} reveals a nice similarity in such a sense that the LVD in figure \ref{mini-cmz} is a superposition of ordinary arms as horizontal LV ridges and largely tilted LV ridges due to the rotating inner disc of high velocity. 
We call the high-velocity region at $|l-l_{\rm SgrA^*}|\lesssim 0\degp07$ the "mini-CMZ". 

Panel C shows a close-up of the mini-CMZ, highlighting the elliptical structure marked by the white ellipse that traces the eastern (E) and western (W) arcs. 
Some well-known molecular features are marked, including the circumnuclear disc (CND) \citep{2018PASJ...70...85T}, 
20-\kms Cloud (20kmC) \citep{Takekawa17}, 
50-\kms Cloud (50kmC) \citep{2009PASJ...61...29T}, 
high-velocity compact cloud (HVCC=CO 0.02-0.02) \citep{oka+1999}, 
and the 'Tadpole' \citep{2023ApJ...942...46K}. 
Panel D shows the same, but in the \hcnaces\ line, highlighting the LV ridge shaped like an arc with less contamination of low-density gas from outer CMZ structures. 
We also stress that the high-positive velocity wing (see below) is less evident on the \hcnaces\ line than in \cs.

\ss{LV Ellipse fit} 
\label{ssslvellipse}
G0.02 (HVCC \citep{2008PASJ...60..429O}) is the main part of the E Arc, protruding from $(l,\vlsr)\sim (0\degp00,-10 \ \ekms)$ and extending toward high velocity reaching $\sim 170$ \kms\  with the highest intensity near $(l,\vlsr) \sim (0\degp015, 120 \ekms)$. 
Another LV arc is recognized at the western end of the mini-CMZ, marked W Arc on the opposite side with respect to \sgrastar.
The large ellipse in panel C of figure \ref{mini-cmz} approximately traces the E and W arcs, and the faint ridge of a long segment of the western ellipse.
In figure \ref{models} (panel A) we plot the positions of the LV ridges read from panel C of figure \ref{mini-cmz}, which can be fitted by a tilted ellipse expressed by
\be
\vlsr=A~x \pm B \sqrt{1-(\Delta x/a)^2},\label{eqmo}
\ee
where $A=dv/dl=1600$ \kms per degree, $B=100$ \kms, $a=0\degp065$ (9.3 pc), and $\Delta x=l+0\degp046$ with the center offset by $0\degp01=1.4$ pc to the east from \sgrastar. 
The entire ellipse is geometrically equivalent to an LV feature that represents a ring of radius $R=9.3$ pc, which rotates at $A=104 \ekms$ and expands at $B=100 \ekms$.
The E Arc for G0.02 is located $0\degp07=10$ pc from \sgrastar.
The rotation curve in the central 10 pc has been measured only approximately due to uncertainties due to suspected non-circular motions \citep{sofue2013}. So, here we adopt the value for $A$ to represent the rotation velocity, $\Vrot \sim 104$ \kms, at $R\sim 10$ pc.
 
\section{Kinematic tracer of the potential}
\ss{Properties of G0.02}
\label{property} 
\citet{oka+1999} estimate the molecular mass to be $M_{\rm mol}\sim 10^5 \Msun$ from the CO-line luminosity.
The radius of the G0.02 cloud is measured as $r\sim \sqrt{\sigma_l \sigma_b}\sim 0.7$ pc on the moment 0 map (Figure \ref{G0.02simu}).
The full width at half maximum of the peak intensity component is measured to be $2\sigma_v \sim 56$ \kms.
The corresponding Virial mass is then calculated as $M_{\rm vir}\sim 3 r \sigma_v^2/G\sim 4\times10^5 \Msun$.
Adopting a wider extent, \citet{oka+1999} obtain an even higher Virial mass of $\sim 3\times 10^6$ to $10^7 \Msun$.
Therefore, the cloud is not gravitationally bound.
The energy required to accelerate the gas to the observed velocity width is then $E_{\rm K}\gtrsim 10^{52}$ erg.
This led to the idea that G0.02 is disturbed by supernova explosions in the cavity at G+0.005-0.030 \citep{oka+1999}.
In this paper, we propose an alternative model that attributes the large velocity width to the shear of orbital motion of gas streaming in the gravitational potential.

The moment 0 map in figure \ref{mini-cmz} suggests that G0.02 is part of an ellipse elongated in the direction of PA$\sim 74\deg$ with the major and minor radii $a\sim 0\degp07(\sim 10 \epc)$ and $b\sim 0\degp01 (\sim 1.4 \epc)$, respectively, or the axial ratio of $b/a\sim 0.14$.
This suggests that G0.02 is part of a molecular ring with the rotation axis at a position angle of $\sim -16\deg$.
The LV ellipse in figure \ref{models} (panel A-c) is geometrically equivalent to a ring of radius $R=9.3$ pc that rotates at $104 \ekms$ and expands at $100 \ekms$ in the LVD.

\ss{Non-circular motion in the extended-mass potential}
In order to explain these kinematical properties, we simulate the evolution of a molecular cloud orbiting in the gravitational potential in the central stellar bulge.
The cloud is represented by an ensemble of $N$ test particles that are initially distributed in a small radius sphere with $r_{\rm c}=0.1~r_0$ ($\sim 1$ pc) of velocity dispersion $\sigma_v=0.1~v_0$ ($\sim 10$ \kms) centered on $(x,y,z,v_x,v_y,v_z)$ (3D Cartesian coordinates and velocities) in the phase space.
Here, $r_0$ and $v_0$ are the scale radius and rotation velocity of the potential, normalized to 1 and non-dimensionalized.
The time is normalized by the rotational period $P=2\pi r_0/v_0$.
We adopt a potential expressed by 
\be
\Phi=1/2 ~ v_0^2 ~{\rm ln} ~\left[\Sigma (x_i/q_i)^2 \right],\label{eqbar}
\ee 
where $q_x:q_y:q_z$ ($=q$ below) gives the axial ratio of the bar \citep{binney+1991}.
A special case of $q=1:1:1$ represents a spherical potential, producing a flat rotation curve.

The integration was performed for a given initial condition without resonance analysis, so that the motion does not necessarily realize closed resonant orbits.
In the test particle simulations here, the hydrodynamic and MHD effects are not evaluated, so the results apply only to the overall orbital behavior in the gravitational potential.
We also neglect the effect of feedback by nuclear activity, except for the expanding ring model.
The general agreement is that solving the initial value problem does not rule out other models, including those concluded here as unlikely.
The results of calculation are displayed in the rest frame of the Cartesian coordinates because the angular speed of the observer (Sun) is sufficiently slower than that of the system considered.
In figure \ref{models} we compare the simulated results for various cases of initial condition and gravitational potential with the observed moment 0 map, LVD and line profile of G0.02 shown in panel A. 

\ss{Simulation}
\sss{Classification of PVD (LVD)}
\def\I{{\it I}-type}\def\U{{\it U}-type}\def\II{{\it II}-type}\def\O{{\it O}-type}

For convenience in describing the simulated PVDs (LVDs), we first classify the anticipated properties considering the current observations in the edge-on discs of the Galaxy and spiral galaxies in the radio lines.\\
\noindent{\I}: This is an inclined straight ridge in the PVD because of the rigid-body motion of a disc or a circular rotation of a ring. \\
\noindent{\U}: This is typical for an eccentric motion in a potential of an extended-mass distribution and barred potential.\\
\noindent{\II}: The so-called parallelogram is a combination of {\it I}- and {\it U}-types.\\
\noindent{{\it O, C}-type}: This happens for a rotating ring superposed by an expanding motion.

\begin{figure*}    
\begin{center}  \vskip -3mm   
(A-a)\includegraphics[height=2.5cm]{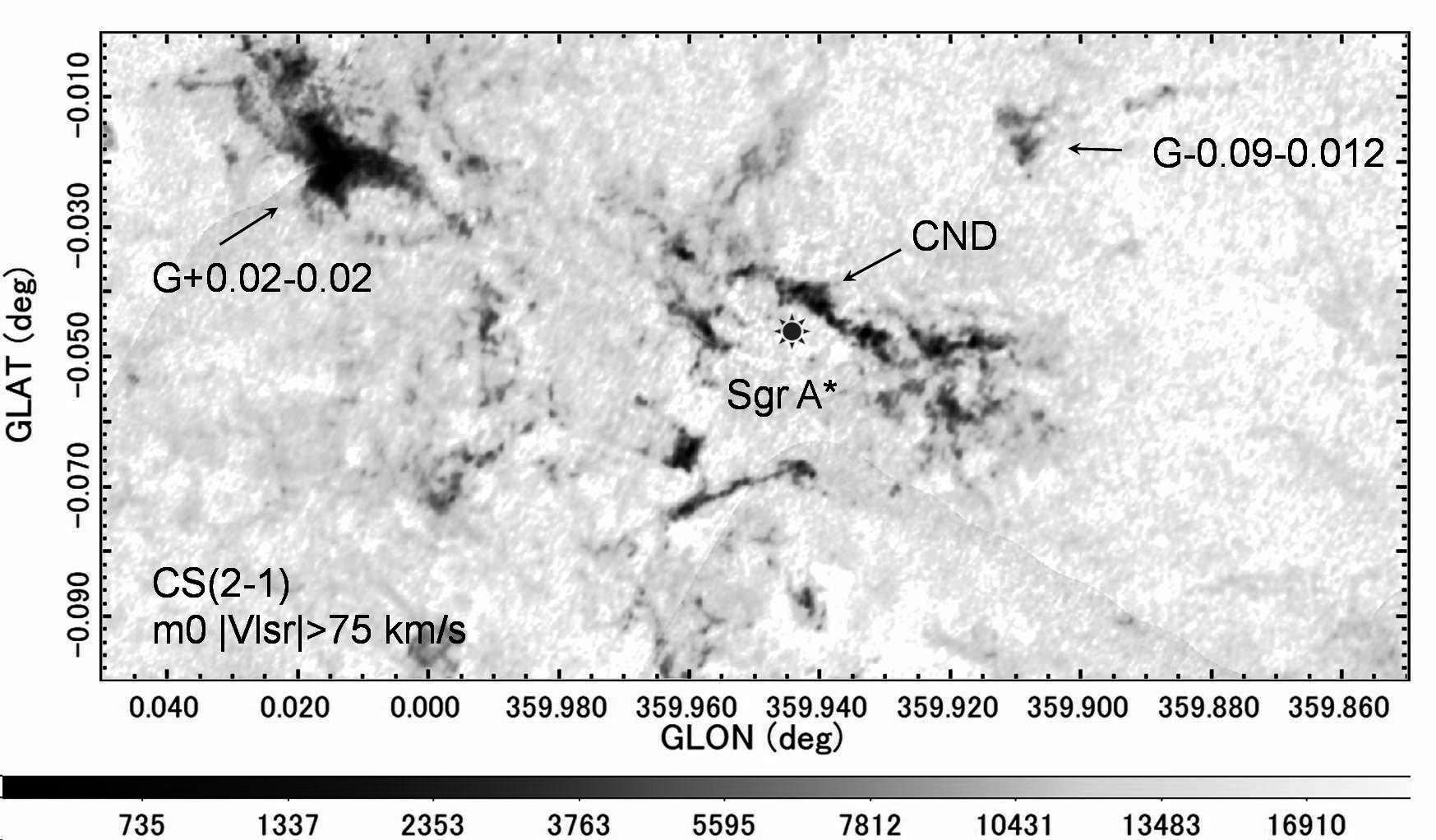}  
(b)\includegraphics[height=2.5cm,width=3.5cm ]{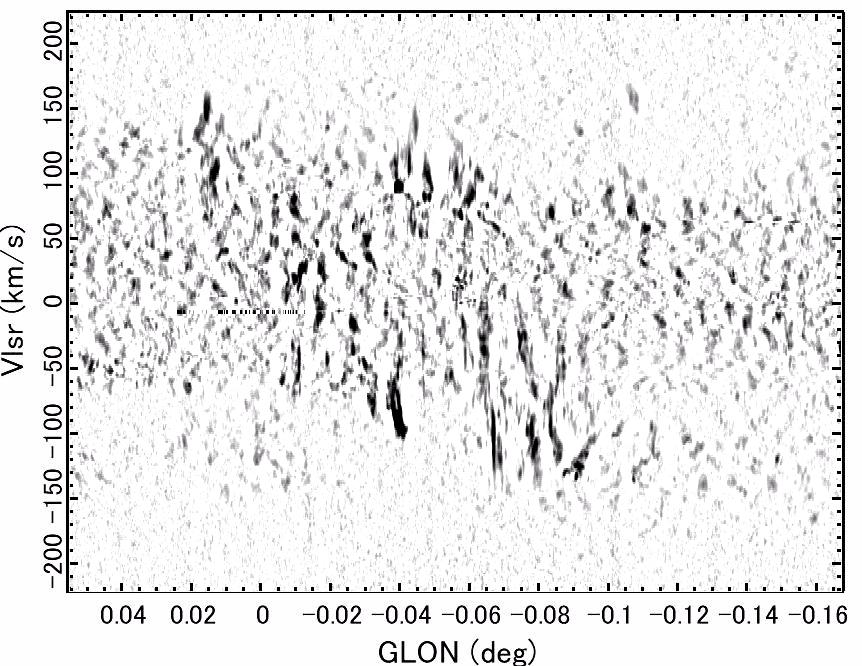} 
(c)\includegraphics[height=2.5cm,width=3.5cm ]{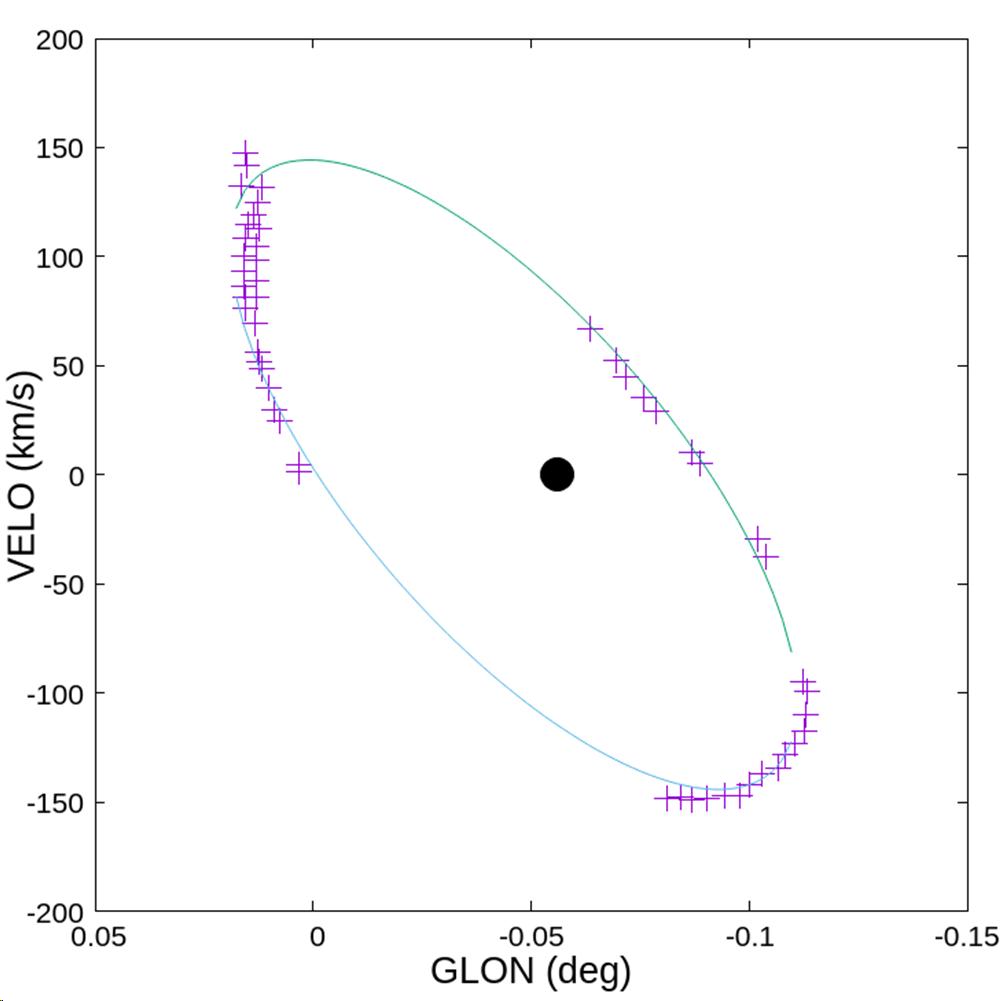} 
(d)\includegraphics[height=2.5cm]{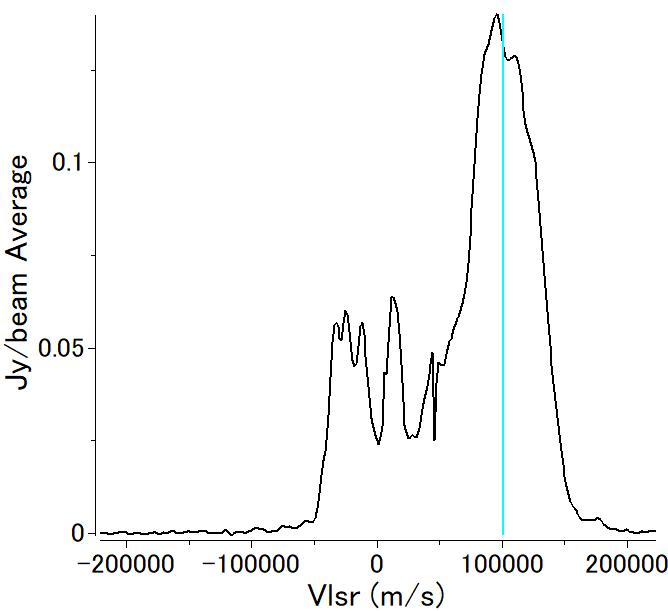}  \\ 
(B)\includegraphics[width=8.2cm]{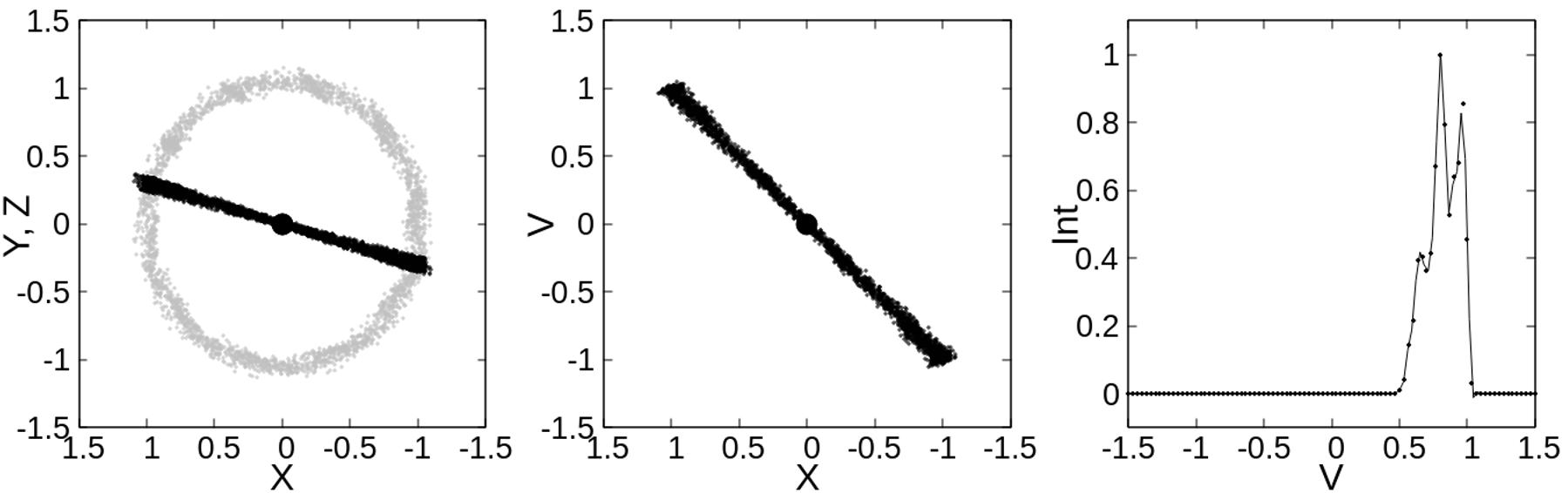}  
(C)\includegraphics[width=8.2cm]{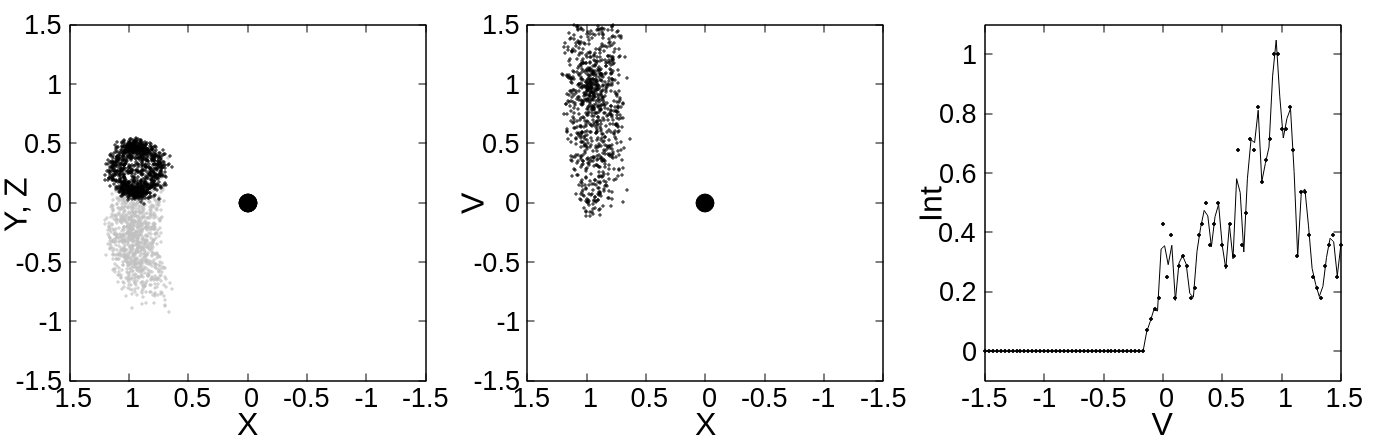} \\   
(D)\includegraphics[width=8.2cm]{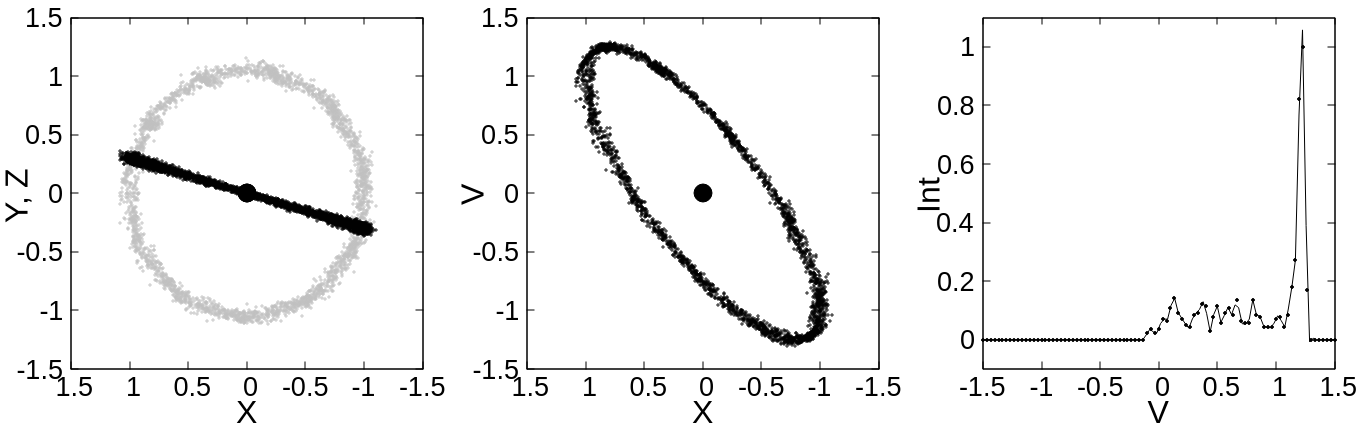}   
(E)\includegraphics[width=8.2cm]{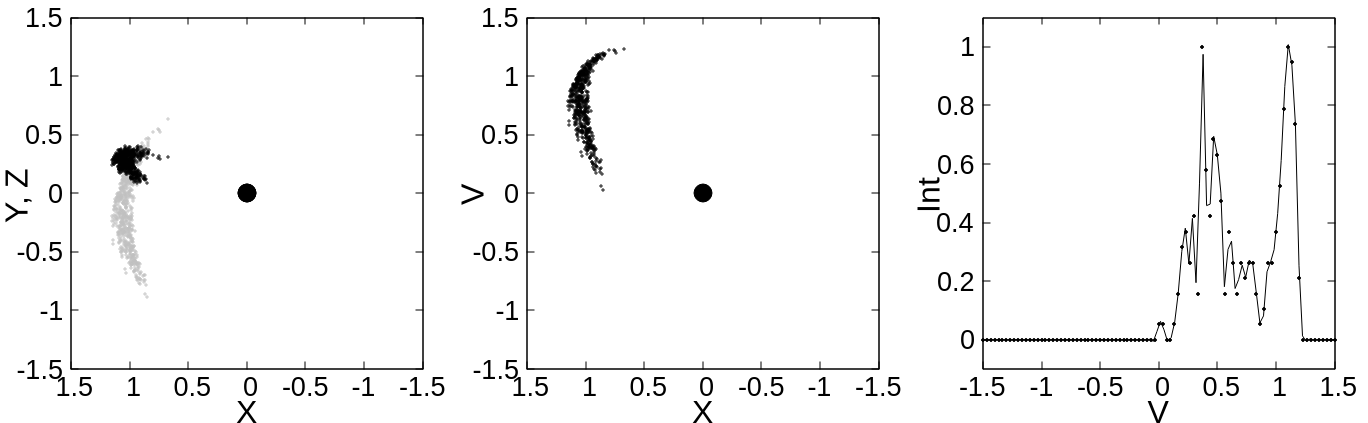} \\ 
(F)\includegraphics[width=8.2cm]{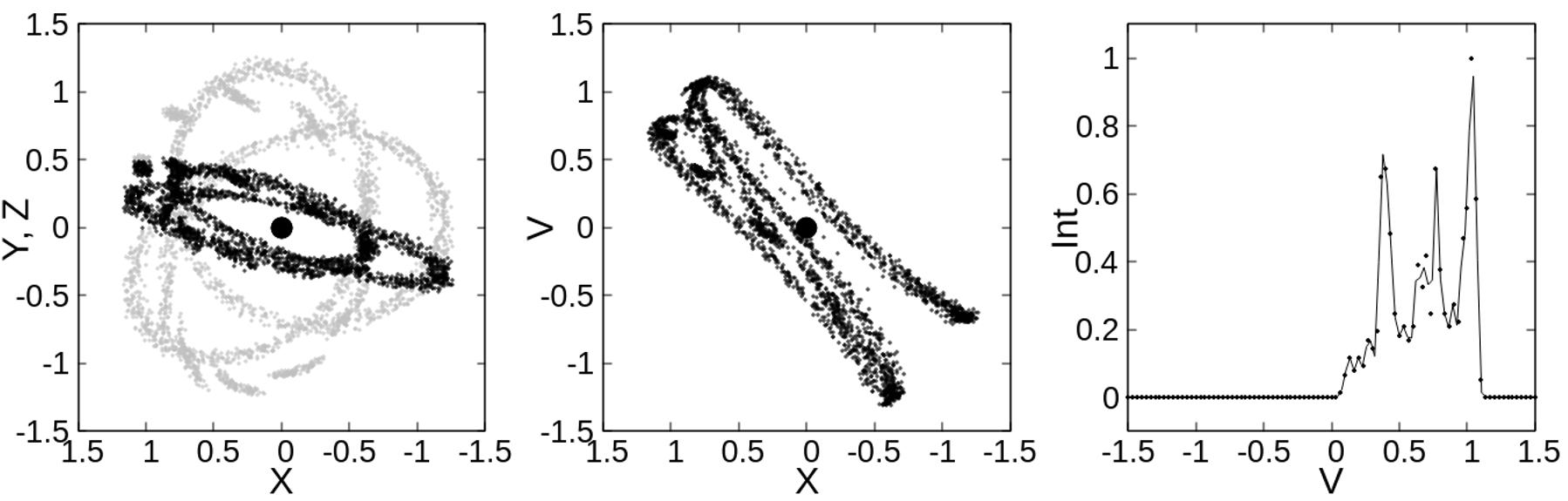}   
(G)\includegraphics[width=8.2cm]{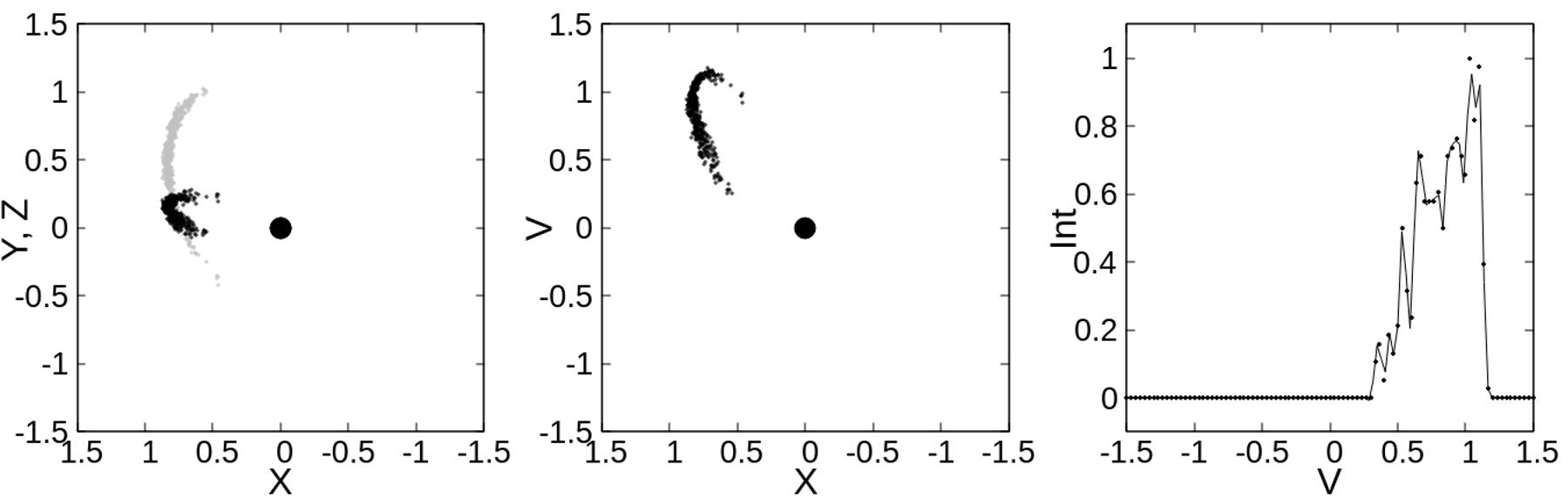}\\
(H)\includegraphics[width=8.2cm]{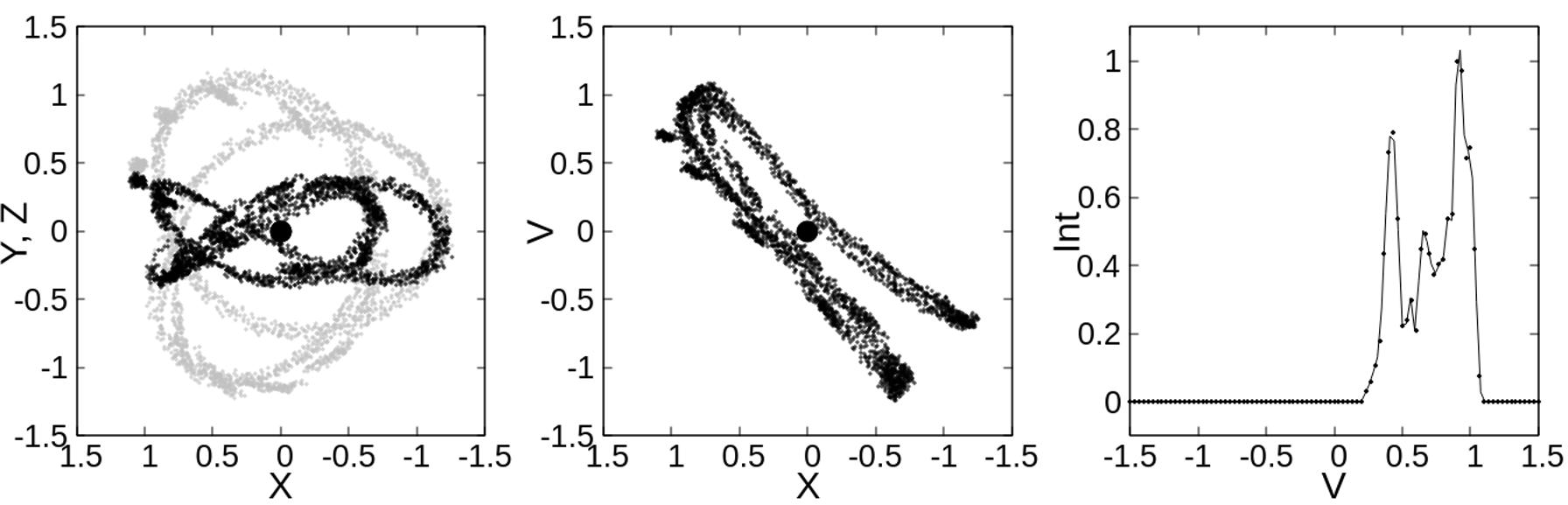}   
(I)\includegraphics[width=8.2cm]{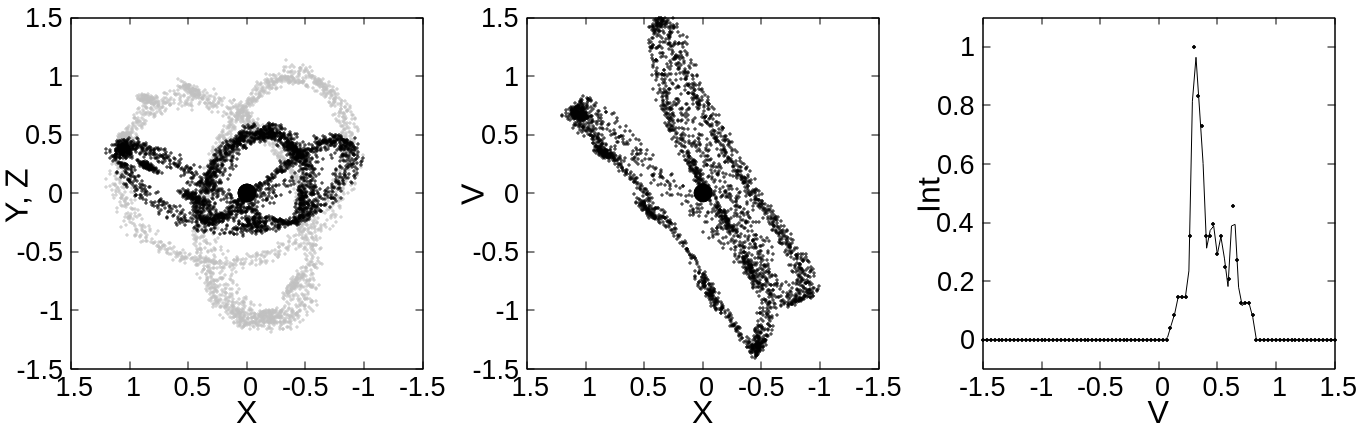}  
\end{center}
\caption{
{Panel [A]}
(a) Observed \cs\-line intensity map at $|\vlsr|\ge 75$ \kms.  
(b) LVD with broad components being subtracted. 
(c) Fitted LV ellipse (Eq. \ref{eqmo}). 
(d) Line profile of G0.02.
{[B]} Circular orbit of a cloud in a spherical potential ($q=1:1:1$) plotted in the $(x,y)$ (gray dots), $(x,z)$ (black), and $(x,v)$ (PVD) planes for initial condition $(\mathbf{r; v})=(1,~0,~0.3; ~0.,~1,~0)$ (in normalized units). 
The 3rd panel shows the line profile at $x\sim 1$. 
Scales are arbitrary.
{[C]} Same, but SN-induced local expansion from the cloud center of $v_{\rm SN}=0.75$ at $t=1$ is added.
{[D]} (\O) Circular flow with radial expansion from the GC at $v_{\rm expa}=0.75$ added, simulating the expanding ring model due to a nuclear activity.
{[E]} Same as D but part at $t\sim 1$, simulating G0.02. 
{[F]} (\U) Eccentric flow in a spherical potential {with initial condition $(\mathbf{r;v})=(1,~0,~0.3;~0.3,~0.8,~0.3)$} plotted every $\delta t=0.1$ from $t=0$ to 3 .
{[G]} (\U) Same, but part at $t\sim 2$, simulating G0.02.
{[H]} ({\U}) Eccentric flow in a disc potential ($q=1:1:0.63$), showing vertical oscillation at a period of $\sim 0.6$ and a $\infty$ shape on the sky.
{[I]} ({\it U},\II) Same, but in a tri-axial bar potential {($q=1:0.85:0.63$)}.
{Note that the wide and lopsided line profile is well reproduced.}
{Alt text: $N$-particle simulations of a cloud orbiting in the gravitational potential in the GC.}
}
\label{models}
\end{figure*}     

\sss{Rotating ring in spherical potential (\I) {- Panel B (figure \ref{models}) -}}
The simplest model is a molecular ring rotating in a spherical potential ($q=1:1:1$) as shown in panel B in figure \ref{models}.
The projection in the $(x,y)$ and $(x,z)$ planes exhibits an ellipse, and the PVD has \I.
However, such a circular rotation is realized only if the gas disc is in hydrostatic equilibrium in the radial direction, which may not be realistic here.
So, in the next subsection, we explore more general orbits in the same potential.

\sss{{SN-induced local expanding cloud {- Panel C -}}}
Since its discovery, G0.02 has been modeled as an expanding molecular shell around a supernova remnant near the center of the arc structure \citep{oka+1999,2008PASJ...60..429O,2022ApJS..261...13O,2023ApJ...950...25I}.
Panel C of figure \ref{models} shows a simulation of such an expanding shell by giving a local radial motion from the cloud's center to each particle.
The result explains the large velocity width, but the line profile is symmetric with respect to the systemic velocity.

\sss{Expanding ring in a spherical potential (\O)  {- Panel D, E -} }
The \O\ PVD has often been explained by an expanding ring \citep{Kaifu+1972,Scoville1972,Sofue2017}.
We examine this model by adding a radial motion to the orbits in the spherical potential.
Figure \ref{models} D shows the result for an expanding motion of $v_{\rm rad}\sim 0.75\, v_0$, and panel E shows a cut at $t\sim 1$.
The radial motion was artificially added to each test particle at a certain epoch when the cloud had reached the final shape. 
This model well reproduces the \O\ LVD observed in the mini-CMZ shown in panel A.
As to the origin of such an expanding motion we may refer to the focusing wave model in a magnetized nuclear gas disc, which postulates the high-efficiency convergence of kinetic energy released at the nucleus to a focal ring at radius $\sim 4.4h$, where $h$ is the scale height of the disc \citep{2020MNRAS.498.1335S}.
The required kinetic energy of expansion of G0.02 is of the order of $E=1/2 M v_{\rm expa}^2\sim 10^{51}$ erg.
Since a few tens of percent of the energy released at the nucleus converges to the cloud, a total energy of $\sim 10^{52}$ of nuclear activity would be sufficient to excite the expanding motion of G0.02. 

\sss{Eccentric motion in a spherical potential (\U) {- Panel F, G -}}
In a spherical potential ($q=1:1:1$) of an extended mass distribution, the orbit of a gas element is generally eccentric and draws a rosette pattern \citep{kruijssen+2015}.
Figure \ref{models} F shows a result of the simulation.
The entire orbit forms a complicated pattern, showing a \U PVD.
However, if we cut part of the orbit as in panel G, it reproduces the observed properties such as the curved arc near the tangential direction, the arc-shaped PVD, and the broad line profile that increases toward the sharp end.
Thus, we conclude that the morphology in the sky and the high-velocity noncircular motion in the PVD of G0.02 are understood to be due to an eccentric orbit of the gas flow in a potential due to an extended mass distribution.
However, in the following, we consider alternative models that may also explain the LV properties.

\sss{Non-circular flow in disc and bar potentials ({\it U}, \II) {- Panel H, I -}}
It is well established that a bar potential produces non-rigid body PVDs such as the parallelogram \citep{binney+1991,kruijssen+2015,sorma+2020}.
Although the bar size of the Milky Way is reported to range from a few to several kpc \citep{2016ARA&A..54..529B,2017MNRAS.465.1621P} and even more inner bars detected in nearby spirals are of sizes $\sim 200$ pc \citep{2024MNRAS.528.3613E}, it would be worth exploring a much smaller bar in the Galactic central bulge.
So, we simulate the evolution of a cloud in a disc and bar of the present size.

Panel H presents the result for an eccentric flow in a flat disc potential with $q=1:1:0.63$ for the initial condition $(\mathbf{r;v})=(1, 0, 0.3; 0.3, 0.8, 0.3)$.
The $(x,y)$ projection forms a rosette orbit, while the $(x,z)$ projection shows a $\infty$ shape, similar to the result for a larger disc of the CMZ \citep{kruijssen+2015}.
The period of vertical oscillation is about equal to $q_z=0.63$ times that of horizontal orbital motion, which is typical of motion in a flat disc.

Panel I shows the result of a cloud orbiting in a triaxial bar potential ($q=1:0.85:0.63$) for the initial condition $(1, 0, 0.3; -0.3, 0.8, 0.3)$.
The LVD behaviors are similar to that in the spherical potential (panel F), and a part of LVD may explain the LVD of G0.02. 

\section{Discussion}
\ss{Summary on the simulated results}
The projected shape in the sky and the global elliptical LVD of G0.02 are generally explained by the eccentric orbit model in the gravitational potential due to an extended mass distribution. 
However, the \O\ LV feature is better reproduced by the expanding-ring model, and the high-velocity LV wing is explained by the SN-induced local expansion model. So, we cannot rule out these models at this time.

\ss{{The Line profile}}
The large velocity width (figure \ref{models} A-d) is a key issue in explaining the origin of the HVCC \citep{oka+1999,2008PASJ...60..429O,2022ApJS..261...13O}, and appears to be ubiquitous in the CMZ \citep{2024ApJ...968L..11G}.
In the right-hand panels of figure \ref{models} we show the simulated line profiles in the tangential direction of the orbit.
The large velocity width is reproduced by the shear of the tangential motion of the elements rotating at high speed in the gravitational potential. 
However, the high-velocity LV wing seems better explained by the SN expansion model but it predicts a more symmetric profile and requires large-energy injection.
Another characteristic of G0.02 is the lopsided line profile shown in figure A-d.
The simulations show that it is a typical profile for the receding side of a rotating disk, exhibiting a sharp cutoff at maximum velocity and an outskirt toward decreasing velocity.

\ss{Origin of G0.02 as a galactic arm}
\sss{Tidal disruption of a cloud}
As simulated, if G0.02 is elongated along the orbit, making it part of a ring or arm, its length is $\sim 16$ pc.
The density of the molecular gas is then estimated to be $n_{\rm H_2}\sim 2\times 10^4$ \htwocc for a total mass of $\sim 10^5\Msun$, much lower than the value of assuming the presence of a spherical cloud. 

\begin{figure}   
\begin{center}   
\includegraphics[width=8.5cm]{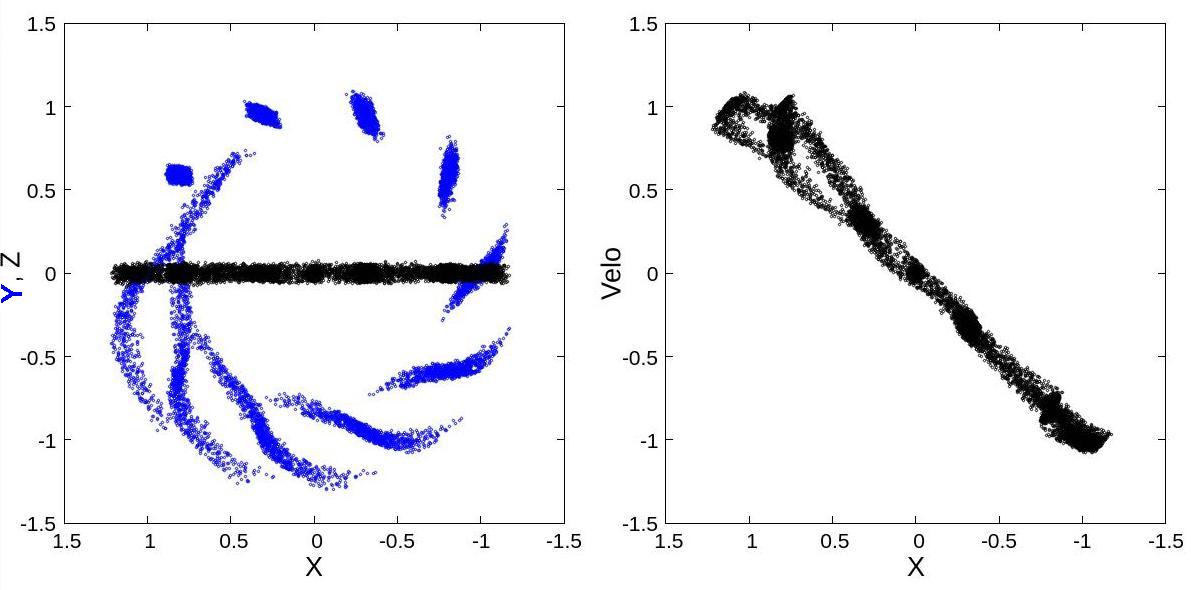} 
 \end{center} 
 \caption{Tidal disruption of a cloud of mass $\sim10^5\Msun$ and radius $\sim1.6$ pc orbiting in the central bulge. {Blue and black dots in the left panel represent projections on $(x,y)$ and $(x,z)$ planes, respectively, at every 0.1 orbital rotation, and the right panel shows LVD $(x,v_y)$. } Molecular clouds can hardly survive in the central bulge.
%Initial condition
%   1.00000005E-03  0.100000001
%   500   1.00   0.10
%    0.00   0.00
%    1.00   1.00   1.00
%    1.00   0.00   0.30   0.00   1.00   0.30   0.00
 {Alt text: Tidal disruption of a cloud.}
 }
\label{fig-tide}
\end{figure}

The concern if it were a cloud would be the tide because of the bulge mass.
For a cloud with mass $m_{\rm c}$ of radius $r$ to survive from tidal disruption, it must be smaller than the Roche radius, $r\lesssim \left( m_{\rm c}/M_{\rm B}\right)^{1/3}R \sim 1.6$ pc, where $M_{\rm B}\sim 2.5\times 10^7\Msun$ is the mass of the bulge inside the orbit $R\sim 10$ pc and $m_{\rm c}\sim 10^5\Msun$ is the mass of the cloud.
Figure \ref{fig-tide} shows an evolution of a cloud represented by test particles distributed around a massive body having the Plummer potential whose center orbits in the same Galactic potential as in figure \ref{models}A. 
The radius of the local potential is $r_{\rm c}\sim1.6$ pc and the mass $m_{\rm _c}\sim10^5\Msun$.
The cloud is strongly stretched to make a spiral shape and loses its identity within one orbital rotation.
Since such a compact cloud can easily be disrupted, it is generally difficult for any molecular cloud to survive disruption in the central 10 pc. 
Therefore, it is more natural to consider that G0.02 is a part of the rotating disc as an arm or a wave orbiting around the potential of the central bulge.

\begin{figure}   
\begin{center}    
\includegraphics[width=5.5cm]{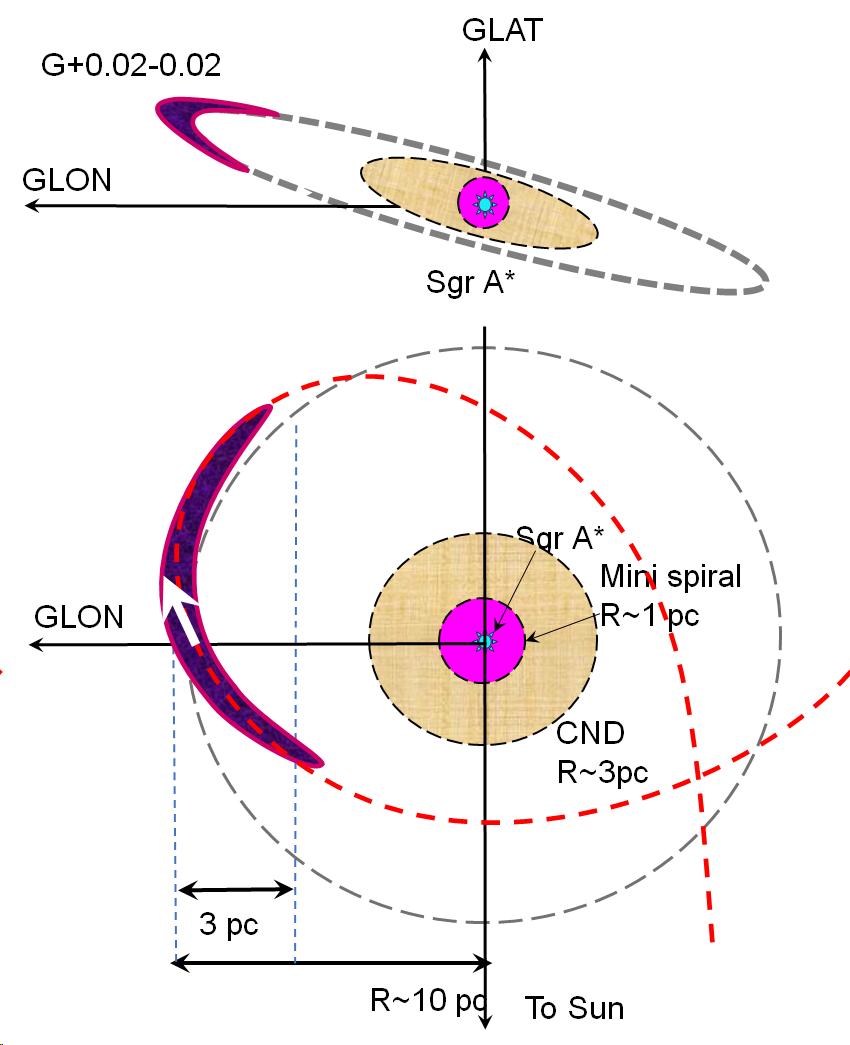} 
 \end{center} 
 \caption{Schematic view of G0.02 and the circum-nuclear gas structures. 
 Top: edge-on view on the sky; 
 bottom: face-on view.
 {Alt text: Illustration of eccentric orbit in the GC simulating G0.02. }
 }
\label{illust}
\end{figure}

\sss{Circum-nuclear arms}
We suggest that G0.02 is an arm that appeared on the circumnuclear gas disc due to the density wave and/or a galactic shock wave, which makes up part of a 10-pc radius tilted ring.
The projection of this ring in the sky forms an ellipse with the major axis at a tilt angle of $\hat{i}\sim 16\deg$ from the Galactic plane ($PA\sim 74\deg$).
This orientation is similar to that of Arm V that has a radius $\sim 8$--9 pc \citep{PaperI} and $\hat{i}\sim 14\deg$ in the same sense.
This coincidence of $\hat{i}$ suggests that G0.02 and Arm V comprise the same family of GC arms.
We point out that the main axes of these features also align with the main axis of the CND \citep{2011AJ....142..134E}.
In Figure \ref{illust} we illustrate a schematic view of the 3D structure in the central 10 pc, where the CND (Arm VI) of the radius $\sim 2.3$ pc and the Mini spirals (Arm VII) of $\sim 1.4$ pc are drawn for comparison.

\ss{Implication for the nuclear kinematics}
G0.02 would provide a new aspect to the study of the circumnuclear region, since it is such a case that a resolved molecular cloud orbits in the deep gravitational potential of the nuclear bulge.
This enables us to explore the 3D potential and therefore the distribution of mass in the inner $\sim 10$ pc region by analyzing the cloud's behavior in the $(l,b,\vlsr)$ space.
This study would be a step toward 3D orbital reconstruction of the gas near Sgr A*, consistent with broader efforts \citep{henshaw+2023,kruijssen+2015}. 
For example, the approximate mass within $R=10$ pc can be estimated using $\vrot\sim 104$ \kms by Eq. \ref{eqmo} as $M=R\vrot^2/G\sim 2.5 \times 10^7 \Msun$ for a spherical mass distribution. 
The 3D visualization of the eccentric gas flow would also help modeling the feeding mechanism to the nucleus.

\ss{Magnetic field aligned along the central arms}
Far infrared polarimetry has shown that the circumnuclear magnetic fields align with the orbital structure in the GC \citep{2025ApJ...982L..22K,2025arXiv250305198Y}.
The magnetic field in G0.02 also aligns exactly with these fields at $PA\sim 70\deg$.
Furthermore, the field within $\sim 2$ pc around G0.02 traces its tightly curved edge \citep{2024MNRAS.531.5012A}.
This magnetic orientation can be naturally understood as a miniature of the galactic magnetic arms, which arise from stretched field lines along the gaseous stream and are compressed in the arm.

%%%%%%%%%%%%%%%%%%%%%%%
\section{Conclusion}
Analyzing the \csaces\ line data cube obtained by ACES, we investigated the detailed kinematical structure of the molecular cloud G0.02.
We showed that G0.02 is an edge-on view of a part of a molecular ring of radius $\sim 10$ pc rotating around \sgrastar\ in $\vrot \sim 100$ \kms superposed by noncircular motion of $\sim 100$ \kms.
The observed properties are consistent with eccentric motion in a gravitational potential of an extended mass distribution. 
Although this is the most natural explanation, expanding ring models can also reproduce LV morphology, particularly \O\ LVD.
Because G0.02 is located near the tangential direction of this ellipse, the width of the molecular line increases to several tens \kms due to shear in orbital velocity.
\G02\ is thus a kinematic tracer of the inner potential, a rare case of a dense gas following an eccentric orbit in the nuclear gravitational field.

%%%%%%%%%%%%%%%%%%%%%%%%%%%%%%%
%\section*{Supplementary data} 
%The following supplementary data is available at PASJ online. 
%%%%%%%%%%%%%%%%%%%% 
{\scriptsize
\begin{ack}
This paper makes use of the following ALMA data: ADS/JAO.ALMA$\#$2021.1.00172: 
/
ALMA is a partnership of ESO (representing its member states), NSF (USA) and NINS (Japan), together with NRC (Canada), NSTC and ASIAA (Taiwan), and KASI (Republic of Korea), in cooperation with the Republic of Chile. 
/
The Joint ALMA Observatory is operated by ESO, AUI/NRAO and NAOJ. 
/
The data analysis in this paper was partially performed at the Astronomical Data Center of the National Astronomical Observatories of Japan.

CB gratefully  acknowledges  funding  from  National  Science  Foundation  under  Award  Nos. 2108938, 2206510, and CAREER 2145689, as well as from the National Aeronautics and Space Administration through the Astrophysics Data Analysis Program under Award ``3-D MC: Mapping Circumnuclear Molecular Clouds from X-ray to Radio,” Grant No. 80NSSC22K1125. 
/
AG acknowledges support from the NSF under AAG 2206511 and CAREER 2142300. 
/
COOL Research DAO \citep{cool_whitepaper} is a Decentralised Autonomous Organisation supporting research in astrophysics aimed at uncovering our cosmic origins. 
/ 
KK acknowledges the support by JSPS KAKENHI Grant Numbers JP23K20035 and JP24H00004. 
/ 
KMD acknowledges support from the European Research Council (ERC) Advanced Grant MOPPEX 833460.vii. 
/
 L.C., V.M.R. and I.J.-S. acknowledge support from the grant PID2022-136814NB-I00 by the Spanish Ministry of Science, Innovation and Universities/State Agency of Research MICIU/AEI/10.13039/501100011033 and by ERDF, UE. 
 /
V.M.R. also acknowledges support from the grant RYC2020-029387-I funded by MICIU/AEI/10.13039/501100011033 and by "ESF, Investing in your future", from the Consejo Superior de Investigaciones Cient{\'i}ficas (CSIC) and the Centro de Astrobiolog{\'i}a (CAB) through the project 20225AT015 (Proyectos intramurales especiales del CSIC); and from the grant CNS2023-144464 funded by MICIU/AEI/10.13039/501100011033 and by “European Union NextGenerationEU/PRTR”. 
/
I.J.-S. acknowledges support from ERC grant OPENS, GA No. 101125858, funded by the European Union. %Views and opinions expressed are however those of the author(s) only and do not necessarily reflect those of the European Union or the European Research Council Executive Agency. Neither the European Union nor the granting authority can be held responsible for them. 
/
P.G. is sponsored by the Chinese Academy of Sciences (CAS), through a grant to the CAS South America Center for Astronomy (CASSACA). 
/  
R.S.K. thanks the 2024/25 Class of Radcliffe Fellows for highly interesting and stimulating discussions, financial support from the European Research Council via  ERC Synergy Grant ``ECOGAL'' (project ID 855130), the German Excellence Strategy via the Heidelberg Cluster of Excellence (EXC 2181 - 390900948) ``STRUCTURES'', the German Ministry for Economic Affairs and Climate Action in project ``MAINN'' (funding ID 50OO2206), Ministry of Science, Research and the Arts of the State of Baden-W\"{u}rttemberg through bwHPC and the German Science Foundation through grants INST 35/1134-1 FUGG and 35/1597-1 FUGG, and also for data storage at SDS@hd funded through grants INST 35/1314-1 FUGG and INST 35/1503-1 FUGG. 
/
D. R-V acknowledges the financial support of DIDULS/ULS, through the project PAAI 2023. 
/
J.W. gratefully acknowledges funding from National Science Foundation under Award Nos. 2108938 and 2206510. 
/
A.S.-M. acknowledges support from the RyC2021-032892-I grant funded by MCIN/AEI/10.13039/501100011033 and by the European Union `Next GenerationEU’/PRTR, as well as the program Unidad de Excelencia María de Maeztu CEX2020-001058-M, and support from the PID2023-146675NB-I00 (MCI-AEI-FEDER, UE). 
/
Q.D.W. acknowledges support from NASA via grant GO3-24120X. 
/
J.K. is supported by the Royal Society under grant number RF\textbackslash ERE\textbackslash231132, as part of project URF\textbackslash R1\textbackslash211322.
/
MCS acknowledges financial support from the European Research Council under the ERC Starting Grant ``GalFlow'' (grant 101116226) and from Fondazione Cariplo under the grant ERC attrattivit\`{a} n. 2023-3014.
/
KMD acknowledges support from the European Research Council (ERC) Advanced Grant MOPPEX 833460.vii.

{The authors thank the anonymous referee for the valuable comments. }
\end{ack} 
 
%\section*{Funding}
 %This research was supported by ...  
 
\section*{Data availability} 
%The single-dish data underlying this article are available at
%https:// www.nro.nao.ac.jp/ $\sim$nro45mrt/html/ results/data.html. 
The interferometer data were taken from the internal release version of the 12m+7m+TP (Total Power)-mode data from the ALMA cycle 8 Large Program "ALMA Central Molecular Zone Exploration Survey" (ACES, 2021.1.00172.L).
% Sample Data Availability Statements 
% https://academic.oup.com/pages/open-research/research-data#Data%20Availability%20Statements
 
\section*{Conflict of interest}
The authors declare that there is no conflict of interest.

} 
%%%%%%%%%%%%%%%%%%%%%%
 

\begin{thebibliography}{}

% Journals(e.g. A\&A,ApJ,AJ,NMRAS,PASP ...)
% Authors, Year, Journal, Vol#, Page#   
%\bibitem[Aauthor et al.(2001)]{key-1}  Aauthor, A., Bauthor, B., \& Cauthor, C.\ 2001, PASJ, 53, 000  

%GC dust polari, mag 2 mG
\bibitem[Akshaya \& Hoang(2024)]{2024MNRAS.531.5012A} Akshaya, M.~S. \& Hoang, T.\ 2024, \mnras, 531, 5012. doi:10.1093/mnras/stae1464    
\bibitem[Binney et al.(1991)]{binney+1991} Binney J., Gerhard O.~E., Stark A.~A., Bally J., Uchida K.~I., 1991, MNRAS, 252, 210   
\bibitem[Bland-Hawthorn \& Gerhard(2016)]{2016ARA&A..54..529B} Bland-Hawthorn, J. \& Gerhard, O.\ 2016, \araa, 54, 529. doi:10.1146/annurev-astro-081915-023441
\bibitem[Chevance et al.(2025)]{cool_whitepaper} Chevance M., Kruijssen J.~M.~D., Longmore S.~N., 2025, doi:10.48550/arXiv.2501.13160
{\bibitem[Etxaluze et al.(2011)]{2011AJ....142..134E} Etxaluze, M., Smith, H.~A., Tolls, V., et al.\ 2011, \aj, The Galactic Center in the Far-infrared, 142, 4, 134. doi:10.1088/0004-6256/142/4/134}
\bibitem[Dame et al.(2001)]{2001ApJ...547..792D} Dame, T.~M., Hartmann, D., \& Thaddeus, P.\ 2001, \apj, 547, 792. doi:10.1086/318388
%Inner 200 pc bar
\bibitem[Erwin(2024)]{2024MNRAS.528.3613E} Erwin, P.\ 2024, \mnras, 528, 3613. doi:10.1093/mnras/stad3944
%UWVeloCloud
\bibitem[Ginsburg et al.(2024)]{2024ApJ...968L..11G} Ginsburg, A., Bally, J., Barnes, A.~T., et al.\ 2024, \apjl, 968, L11. doi:10.3847/2041-8213/ad47fa 
%R0 = 8178 pc: 13stat: 22sys: pc
\bibitem[Gravity Collaboration et al.(2019)]{gravity+2019} Gravity Collaboration, Abuter, R., Amorim, A., et al.\ 2019, \aap, 625, L10. doi:10.1051/0004-6361/201935656 
\bibitem[Henshaw et al.(2023)]{henshaw+2023} Henshaw, J.~D., Barnes, A.~T., Battersby, C., et al.\ 2023, Protostars and Planets VII, 534, 83. doi:10.48550/arXiv.2203.11223
%\bibitem[Heywood et al.(2022)]{2022ApJ...925..165H} Heywood, I., Rammala, I., Camilo, F., et al.\ 2022, \apj, 925, 165. doi:10.3847/1538-4357/ac449a
%CND ALMA
\bibitem[Hsieh et al.(2021)]{2021ApJ...913...94H} Hsieh, P.-Y., Koch, P.~M., Kim, W.-T., et al.\ 2021, \apj, 913, 94. doi:10.3847/1538-4357/abf4cd  
%hvcc
\bibitem[Iwata et al.(2023)]{2023ApJ...950...25I} Iwata, Y., Oka, T., Takekawa, S., et al.\ 2023, \apj, 950, 25. doi:10.3847/1538-4357/acc9b0
\bibitem[Kaifu et al.(1972)]{Kaifu+1972} Kaifu, N., Kato, T., \& Iguchi, T.\ 1972, Nature Physical Science, 238, 105 
%Tadpole
\bibitem[Kaneko et al.(2023)]{2023ApJ...942...46K} Kaneko, M., Oka, T., Yokozuka, H., et al.\ 2023, \apj, 942, 46. doi:10.3847/1538-4357/aca66a
%B field // G0.02 ring
\bibitem[Karoly et al.(2025)]{2025ApJ...982L..22K} Karoly, J., Ward-Thompson, D., Pattle, K., et al.\ 2025, \apjl, %The JCMT BISTRO Survey: Magnetic Fields Align with Orbital Structure in the Galactic Center, 
982, 1, L22. doi:10.3847/2041-8213/adbc67
\bibitem[Kruijssen et al.(2015)]{kruijssen+2015} Kruijssen, J.~M.~D., Dale, J.~E., \& Longmore, S.~N.\ 2015, \mnras, 447, 1059. doi:10.1093/mnras/stu2526  
%hvcc +0.02-0.02
\bibitem[Oka et al.(2008)]{2008PASJ...60..429O} Oka, T., Hasegawa, T., White, G.~J., et al.\ 2008, \pasj, 60, 429. doi:10.1093/pasj/60.3.429% new look at Sgr A in mol line
%high velo cloud 
\bibitem[Oka et al.(2022)]{2022ApJS..261...13O} Oka, T., Uruno, A., Enokiya, R., et al.\ 2022, \apjs, 261, 13. doi:10.3847/1538-4365/ac6bfc
%hvcc
\bibitem[Oka et al.(1999)]{oka+1999} Oka, T., White, G.~J., Hasegawa, T., et al.\ 1999, \apj, 515, 249. doi:10.1086/307029  
%MW Bar
\bibitem[Portail et al.(2017)]{2017MNRAS.465.1621P} Portail, M., Gerhard, O., Wegg, C., et al.\ 2017, \mnras, 465, 1621. doi:10.1093/mnras/stw2819
\bibitem[Scoville (1972)]{Scoville1972} Scoville, N.~Z.\ 1972, \apjl, 175, L127 
\bibitem[Sofue(2013)]{sofue2013} Sofue, Y.\ 2013, \pasj, 65, 118. doi:10.1093/pasj/65.6.118%GC RC 
\bibitem[Sofue(2017)]{Sofue2017} Sofue, Y.\ 2017, \mnras, 470, 1982%200pc 
%mhd feedback sgr a to b
\bibitem[Sofue(2020)]{2020MNRAS.498.1335S} Sofue, Y.\ 2020, \mnras, 498, 1335. doi:10.1093/mnras/staa2389  
%\bibitem[Sofue et al.(1986)]{1986ARA&A..24..459S} Sofue, Y., Fujimoto, M., \& Wielebinski, R.\ 1986, \araa, %Global structure of magnetic fields in spiral galaxies., 24, 459. doi:10.1146/annurev.aa.24.090186.002331
%Paper I ACES
\bibitem[Sofue et al.(2025)]{PaperI} Sofue, Y. et al. 2025, PASJ in press, arXiv:2504.03331 (Paper I)
\bibitem[Sormani et al.(2020)]{sorma+2020} Sormani, M.~C., Tress, R.~G., Glover, S.~C.~O., et al.\ 2020, \mnras, 497, 5024.  doi:10.1093/mnras/staa1999 
%20kmC
\bibitem[Takekawa et al.(2017)]{Takekawa17} Takekawa, S., Oka, T., \&  Tanaka, K.\ 2017, \apjl, 834, 121. doi:10.3847/1538-4357/834/2/121
%CND alma
\bibitem[Tsuboi et al.(2018)]{2018PASJ...70...85T} Tsuboi, M., Kitamura, Y., Uehara, K., et al.\ 2018, \pasj, 70, 85. doi:10.1093/pasj/psy080
%50kmc
\bibitem[Tsuboi et al.(2009)]{2009PASJ...61...29T} Tsuboi, M., Miyazaki, A., \& Okumura, S.~K.\ 2009, \pasj, 61, 29. doi:10.1093/pasj/61.1.29
\bibitem[Yang et al.(2025)]{2025arXiv250305198Y} Yang, M.-Z., Lai, S.-P., Karoly, J., et al.\ 2025, 
%, The JCMT BISTRO Survey: Unveiling the Magnetic Fields around Galactic Center, 
arXiv:2503.05198. doi:10.48550/arXiv.2503.05198
\end{thebibliography}
\end{document}